\documentclass{imsart}

\RequirePackage[OT1]{fontenc}
\RequirePackage{amsthm,amsmath,amssymb}
\RequirePackage[numbers]{natbib}
\RequirePackage{hyperref}
\RequirePackage{graphics}
\RequirePackage{sidecap}
\RequirePackage{fancybox}
\RequirePackage{tikz}
\RequirePackage{wrapfig}
\RequirePackage{multicol}
\RequirePackage{multirow}
\RequirePackage{graphicx}
\RequirePackage{subcaption}

\def\iid{\stackrel{\text{iid}}{\sim}}
\def\argmin{\mathop{\rm argmin}}
\def\det{\mbox{det}}
\def\tr{\mbox{tr}}

\def\i{{\bf i}}
\def\r{{\bf r}}

\def\M{{\cal M}}

\def\si2{\sigma^2}
\def\ep{\epsilon}
\def\om{\omega}

\def\de{\delta}
\def\Si{\Sigma}
\def\la{\lambda}

\def\R{\mathbb{R}}

\def\iid{\overset{iid}{\sim } }
\def\Jsum{\sum_{\substack{\i=(i_1,\cdots,i_{p}) \\ 1\leq i_1<\cdots<i_{p}\leq n}}}

\def\Jusum{\sum_{\substack{\i=(\i_1, \cdots, \i_p)\\ \forall i,\; \i_i=(i_{i,1},\cdots,i_{i,p_i}),\\1\leq i_{i,1}<\cdots<i_{i,p_i}\leq n}}}
\def\Jsumgi{\sum_{\substack{\i_i=(i_1,\cdots,i_{g_i}) \\ 1\leq i_1<\cdots<i_{g_i}\leq n}}}

\def\dWA{\frac{\partial \textbf{w}}{\partial A}}
\def\d{{\bf d}}
\def\W{{\bf w}}

\def\E{\mbox{E}}

\usepackage{pdfsync}

\newcommand{\toP}{\overset{P}{\longrightarrow}}
\newcommand{\toD}{\overset{\mathcal D}{\longrightarrow}}

\newcommand{\sbul}{\lower 3pt\hbox{\LARGE $\cdot\,$}}

\newcommand{\cM}{{\mathcal{M}}}

\newcommand{\bY}{{\boldsymbol{Y}}}
\newcommand{\by}{{\boldsymbol{y}}}
\newcommand{\bZ}{{\boldsymbol{Z}}}

\def\half{\frac{1}{2}}

\definecolor{shadecolor}{gray}{.65}

\startlocaldefs
\numberwithin{equation}{section}
\theoremstyle{plain}
\newtheorem{thm}{Theorem}[section]
\newtheorem*{thm*}{Theorem}
\newtheorem{prop}{Proposition}[section]
\newtheorem*{prop*}{Proposition}
\newtheorem{lemma}{Lemma}[section]
\newtheorem*{lemma*}{Lemma}

\newtheorem{defi}{Definition}[section]

\endlocaldefs

\begin{document}

\begin{frontmatter}
\title{Covariance estimation via fiducial inference}
\runtitle{Fiducial covariance estimation}

\begin{aug}
\author{\fnms{W. Jenny} \snm{Shi}\thanksref{t1}\corref{}\ead[label=e1]{wjennyshi@gmail.com}}
\address{Computational Bioscience Program, University of Colorado
	\printead{e1}}
\author{\fnms{Jan} \snm{Hannig}\thanksref{t2}
\ead[label=e2]{jan.hannig@unc.edu}}
\address{Department of Statistics \& Operations Research, University of North Carolina}
\author{\fnms{Randy C.S.} \snm{Lai}
\ead[label=e3]{chushing.lai@maine.edu}}
\address{Department of Mathematics \& Statistics, University of Maine
	\printead{e3}}
\author{\fnms{Thomas C.M.} \snm{Lee}\thanksref{t3}
\ead[label=e4]{tcmlee@ucdavis.edu}}
\address{Department of Statistics, University of California
	\printead{e4}}

\thankstext{t1}{Shi's research was supported in part by the National Library of Medicine Institutional Training Grant T15LM009451}
\thankstext{t2}{Hannig's research was supported in part by the National Science Foundation (NSF) under Grant  No.\ 1512945 and 1633074}
\thankstext{t3}{Lee's research was supported in part by the NSF under Grant  No.\ 1512945 and 1513484}
\runauthor{Shi, Hannig, Lai, Lee}

\end{aug}

\begin{abstract}
As a classical problem, covariance estimation has drawn much attention from the statistical community for decades. Much work has been done under the frequentist and the Bayesian frameworks. Aiming to quantify the uncertainty of the estimators without having to choose a prior, we have developed a fiducial approach to the estimation of covariance matrix. Built upon the Fiducial Berstein-von Mises Theorem (Sonderegger and Hannig 2014), we show that the fiducial distribution of the covariate matrix is consistent under our framework. Consequently, the samples generated from this fiducial distribution are good estimators to the true covariance matrix, which enable us to define a meaningful confidence region for the covariance matrix. Lastly, we also show that the fiducial approach can be a powerful tool for identifying clique structures in covariance matrices. 

\end{abstract}



\end{frontmatter}

\section{Introduction}
Estimating covariance matrices has always been an important task. Many regression-based methods have emerged in the last few decades, especially in the concept of ``large $p$ small $n$''. Among the notable methods, there are the graphical LASSO algorithms \cite{Friedman2008, Friedman2010, Rothman2012}. Pourahmadi provided a detailed overview on the progress of covariance estimation \cite{Pourahmadi2011}. The Positive Definite Sparse Covariance Estimators (PDSCE) method \cite{Rothman2012} has grained great popularity due to its performance comparing to other current methods, although it only produces a point estimator. 

Aiming to have a distribution of ``good'' covariance estimators, we propose a generalized fiducial approach. The general approach of fiducial inference was first proposed by Ronald A. Fisher in the 1930's, whose intention was to overcome the need for priors and other problems with Bayesian methods at the time. The procedure of fiducial inference allows to obtain a measure on the parameter space without requiring priors and defines approximate pivots for parameters of interest. It is ideal when {\it a priori} information about the parameters is unavailable. The key recipe of the fiducial argument is the data generating equation. Roughly, the generalized fiducial likelihood is defined as the distribution of the functional inverse of the data generating mechanism. 

One great advantage of the fiducial approach to covariance matrix estimation is that, without specifying a prior, it produces a family of matrices that are close to the true covariance with a probabilistic characterization using the fiducial likelihood function. This attractive property enables a meaningful definition for matrix confidence regions. 

We are particularly interested in a high-dimensional multivariate linear model setting with possibly an atypical sparsity constraint. Instead of classical sparsity assumptions on the covariance matrix, we consider a type of experimental design that enforces sparsity on the covariate matrix. This phenomenon often arises in the studies of metabolomics and proteomics. One example of this setup is modeling the relationship between a set of gene expression levels and a list of metabolomic data. The expression levels of the genes serve as the predictor variables while the response variables are a variety of metabolite levels, such as sugar and triglycerides. It is known that only a small subset of genes contribute to each metabolite level, and each gene can be responsible for just a few metabolite levels.  

Under the sparse covariate setting, we derive the generalized fiducial likelihood of the covariate matrix based on given observations and prove its asymptotic consistency as the sample size increases. For the covariance with community structures (cliques), we prove the necessary conditions for achieving accurate clique structure estimation. Samples from the fiducial distribution of a covariate matrix can be generated using Monte Carlo methods. In the general case, a reversible jump Markov chain Monte Carlo (RJMCMC) algorithm may be needed. Similar to the classic likelihood functions, fiducial distributions favor models with more parameters. Therefore, in the case where the exact sparsity structure of the covariate is unclear, a penalty term needs to be added. To obtain a family of covariance estimators in the general case, we adapt a zeroth-order method and develop an efficient RJMCMC algorithm that samples from the penalized fiducial distribution.

The rest of the paper is arranged as follows. In Section~\ref{sec:GFI} we will provide a brief background and development on fiducial inference. Then we will introduce the fiducial model for covariance estimation and derive the Generalized Fiducial Distribution (GFD) for the covariate and covariance matrices in Section~\ref{sec:COVmodel}. We will examine the asymptotic property of the GFD of the covariance matrix under minor assumption, and show that it satisfies the Fiducial Bernstein von-Mises Theorem in Section~\ref{sec:theory}.  Section~\ref{sec: ModelSelection} focuses on model selection, where we show some theoretical results for the clique model and how to choose appropriate penalty terms.  In Section~\ref{sec:simulation} we examine the samples from the GFD's, including both the clique models and the general case. Additional simulations can be found in the supplementary.  Finally, Section~\ref{sec:COVdiscussion} concludes the paper with a summary and a short discussion on the relationship of our approach to Bayesian methods.

\section{Generalized fiducial inference}\label{sec:GFI}

\subsection{Brief background}
Fiducial inference was first proposed by Ronald Aylmer Fisher when he introduced the concept of a fiducial distribution of a parameter in 1930. In the case of a single parameter family of distributions, Fisher gave the following definition for a {\it fiducial density} $f(\theta|x)$ of the parameter based on a single observation $x$ for the case where the cumulative distribution function $F(x|\theta)$ is a monotonic decreasing function of $\theta$:
\begin{equation}\label{eq:Fisher}
f(\theta|x)\propto -\frac{\partial F(x|\theta)}{\partial \theta}
\end{equation}
 A fiducial distribution can be viewed as a Bayesian posterior distribution without hand picking priors. In many single parameter distribution families, Fisher's fiducial intervals coincide with classical confidence interval. For families of distributions with multiple parameters, the fiducial approach leads to confidence set. The definition of fiducial inference has been generalized in the past decades. \cite{Hannig2015} provides a detailed review on the philosophy and current development on the subject.
 
 The generalized fiducial approach has been applied to a variety of models, both parametric and nonparametric, both continuous and discrete. These applications include bioequivalence \cite{Hannig2006b}, variance components \cite{Cisewski2012,E2008}, problems of metrology \cite{Hannig2007,Hannig2003,Wang2012,Wang2005,Wang2006,Wang2006b}, inter laboratory experiments and international key comparison experiments \cite{Iyer2004}, maximum mean of a multivariate normal distribution \cite{Wandler2011}, multiple comparisons \cite{Wandler2012}, extreme value estimation \cite{Wandler2012b}, mixture of normal and Cauchy distributions \cite{Glagovskiy2006}, wavelet regression \cite{Hannig2009b}, logistic regression and LD$_{50}$ \cite{Epreprint}.\\

\subsection{Generalized fiducial distribution} 
The idea underlying generalized fiducial inference is built upon a {\it data generating equation} $G(\cdot,\cdot)$ expressing the relationship between the data $X$ and the parameters $\theta$:
\begin{equation}\label{eq:G}
X=G(U,\theta),
\end{equation}
where $U$ is the random component of this data generating equation whose distribution is known. The data $X$ are assumed to be created by generating a random variable $U$ and plugging it into the date generating equation above.

The generalized fiducial distribution (GFD) inverts Eq \ref{eq:G}. Assume that $x\in \R^n$ is continuous, and the parameter $\theta\in\R^p$. Under the conditions provided in \cite{Hannig2015}, the fiducial density is
        \begin{equation}\label{eq:r}
        r(\theta|x)=\frac{f(x,\theta)J(x,\theta)}{\int_{\Theta}f(x,\theta')J(x,\theta')d\theta'},
        \end{equation}
        
where $f(x,\theta)$ is the likelihood; $J(x,\theta)$ is the Jacobian function:
\begin{equation}\label{eq:Jacob2}
J(x,\theta)=D \left(\left. \frac{d}{d\theta}G(u,\theta)\right|_{u=G^{-1}(x,\theta)}\right)
\end{equation}

Exact form of $D(\cdot)$ depends on the choices of norm in the inverting procedure. In the remaining part of this paper, we will focus on the $l_2$- and the $l_\infty$-norms. Under the $l_2$-norm, 
\begin{equation}\label{eq:L2D}
        D(M) = \sqrt{\det(M^TM/n)};
\end{equation}

when the $l_\infty$-norm is used, 
\begin{equation}\label{eq:LinftyD}
D(M) = \begin{pmatrix}n\\p \end{pmatrix}^{-1}\Jsum\left|\det \left(M\right)_\i \right|.
\end{equation}

For any matrix $M$ of size $n\times p$, $n\geq p$, $(M)_\i$ stands for a $p\times p$ matrix consisting of rows $\i=(i_1,\cdots,i_p)$ of $M$. 
The summation above goes over all $p$-tuples of indexes $\i=(1\leq i_1<\cdots<i_p\leq n)\subset\{1,\cdots,n\}$; $\det \left(M\right)_\i$ is the determinant of the sub-matrix $M_\i$.

\section{A fiducial approach to covariance estimation}{\label{sec:COVmodel}}
In this section we will derive the generalized fiducial distriubtion (GFD) for the covariance matrix of a multivariate normal random variable. Let $Q^T$ denote the transpose of a matrix/vector $Q$. Denote a collection of $n$ observed $p$ dimensional objects $\textbf{Y}=\{Y_i,\;i=1,\cdots,n\}$. For the rest of the paper, we assume $p$ is fixed, unless stated otherwise. Consider the following data generating equation: 
\begin{equation}\label{eq:model}
Y_i=AZ_i,\;i=1,\cdots,n;
\end{equation}
where $A$ is a $p\times p$ matrix of full rank; $\bZ=\{Z_i=(z_{i1},\cdots, z_{ip})^T,\;i=1,\cdots,n\}$ are independent and identically distributed ({\it i.i.d}) $p\times 1$ random vectors following multivariate normal distribution $N(0,I)$. Hence, $Y_i$'s are {\it i.i.d} random vectors centered at 0 with variance $AA^T$,
\begin{equation}\label{eq:Yi}
\text{ i.e. }Y_i\iid N(0,\Sigma),\text{ where }\Sigma=AA^T. 
\end{equation}
Consequently, we have the likelihood for observations $\by$:
\begin{equation}\label{eq:normal}
f(\by,A)=(2\pi)^{-\frac{np}{2}}|\det(A)|^{-n}\exp\left[-\frac{1}{2}\tr\{nS_n(AA^T)^{-1}\}  \right],
\end{equation}
where $Sn=\frac{1}{n}\sum_{i=1}^ny_iy_i^T$ is the corresponding sample covariance matrix and $tr\{\cdot\}$ is the trace operator.

We propose to estimate the covariance matrix $\Sigma$ through the GFD of covariate matrix $A$:
\begin{equation}\label{eq:GFDA}
r(A| \by)\propto J(\by,A)f(\by,A)
\end{equation}
 
Define the stacked observation vector $\W = (y_1^T, \cdots, y_n^T)^T = (w_1, \cdots, w_{np})^T$. Denote ${\bf u} = (u_1, \cdots, u_n)$, such that $y_i = G(u_i, A),\;\forall i$. Let $a_{ij}$ be the $ij$th entry of matrix $A$, i. e. $A=[a_{ij}]_{1\leq i,j\leq p}$. The corresponding Jacobian $J(\by,A)$ derived from (\ref{eq:Jacob2}) is then
 \begin{equation}\label{eq:bigJ}
J(\by,A)=D\left(\left.\det\left(\dWA\right)\right|_{{\bf u}=G^{-1}(\by,A)}\right),
\end{equation}
and $\dWA$ is an $np\times p^2$ matrix:
\begin{equation*}
\dWA=
\begin{pmatrix}
\frac{\partial w_{1}}{\partial a_{11}} & \frac{\partial w_{1}}{\partial a_{12}} & \cdots & \frac{\partial w_{1}}{\partial a_{pp}}\\
\frac{\partial w_{2}}{\partial a_{11}} & \frac{\partial w_{2}}{\partial a_{12}} & \cdots & \frac{\partial w_{2}}{\partial a_{pp}}\\
\vdots & \vdots & \ddots & \vdots\\
\frac{\partial w_{np}}{\partial a_{11}} & \frac{\partial w_{np}}{\partial a_{12}} & \cdots & \frac{\partial w_{np}}{\partial a_{pp}}
\end{pmatrix}.
\end{equation*}

Note that if $a_{kl} $ is fixed at zero, then the $kl$th column vanishes. Depending on the parameter space, the dimension and the exact form of $\dWA$ varies.

\subsection{Jacobian for full models}\label{Sec:FullModelJacobian}

Suppose that none of the entries of $A$ is fixed at zero, namely, the parameter space $\Theta$ for $A$ is $\R^{p\times p}$. We will refer this to a full model. Under a full model, $\dWA$ consists of $p$ blocks, each of dimension $np\times p$. Every row of $\dWA$ has non-zero entries in only one block.

By swapping rows in the matrix $\dWA$ and plugging ${\bf u}=G^{-1}(\by,A)$, we obtain the $np\times p^2$ matrix $P$:
\begin{equation}\label{eq:P}
P
= \begin{pmatrix}
U &  &  \\
 & \ddots & \\
 & & U
\end{pmatrix}
\end{equation}
where $U = (A^{-1}y_{1}; \cdots; A^{-1}y_{n})^T = V(A^{-1})^T, V =(y_{1};\cdots;y_{n})^T.$ $P$ breaks into $p$ blocks, $B_1,\cdots, B_p$, where $B_i = \begin{pmatrix}
O_{(ni-n)\times p}\\ U \\ O_{(np- ni) \times p}
\end{pmatrix}$, $O_{a\times b}$ denotes a zero matrix with dimension $a\times b$.

Since swapping rows does not change the absolute value of the determinant of a matrix, the Jacobian (\ref{eq:bigJ}) can be expressed with matrix $P$:
\begin{equation}\label{eq:fullAJ}
J(\by,A)=  D\left(\det\left(P\right)\right)=C(\by)|\det(A)|^{-p}.
\end{equation}

where the Jacobian constant 
\begin{equation}\label{eq:CJ}
C(\by) = \begin{cases}
\left|\det(S_n)\right|^\frac{p}{2}, & \text{ under } l_2 \text{ norm;}\\
\begin{pmatrix}n\\p \end{pmatrix}^{-p}\left(\Jsum\left|\det\left(V\right)_\i \right|\right)^p, & \text{ under } l_\infty \text{ norm.}
\end{cases}
\end{equation}

By (\ref{eq:GFDA}), the GFD is proportional to 
\begin{equation}\label{eq:fullGFD}
r(A|\by)\propto C(\by)(2\pi)^{-\frac{np}{2}}|\det(A)|^{-(n+p)}\exp\left[-\frac{1}{2}\tr\{nS_n(AA^T)^{-1}\}  \right].
\end{equation}
By transforming the GFD of $A$, we conclude that the GFD of $\Sigma=AA^T$ has the inverse Wishart distribution with $n$ degrees of freedom and parameter $nS_n.$  \\

\subsection{Jacobian for clique model}

While having a closed form for the GFD of $\Sigma$, the covariance estimation requires sufficient number of observations (roughly at least $n > 15(p + 1)$) to maintain reasonable power. In the cases where n is small, we ``reduce'' the parameter space by introducing a sparse structure $\cM$, which defines the parameter space for $A$. The full model is a special case where $\cM$ corresponds to $\R^{p\times p}$.

Assume that the coordinates of $\by$ are broken into cliques; i.e. coordinates $i$ and $j$ are correlated if $i,j$ belong to the same clique and independent otherwise. By simply swapping rows and columns of the covariate matrix, we can arrive at a block diagonal form. Without loss of generality, suppose that $A$ is a block diagonal matrix with block sizes $g_1, \cdots, g_k$. Then its model $\cM$ defines the parameter space $\otimes_i^k \R^{g_i \times g_i}$. Given $\cM$, as an extension of the full model, the GFD function in this case becomes a composite of inverse Wishart distributions:

\begin{equation}\label{eq:IWclique}
r(\Sigma | \by, \cM) = \prod_i^k \frac{|nS_n^i|^{\frac{n}{2}}}{2^\frac{ng_i}{2} \Gamma_{g_i}\left(\frac{n}{2} \right)} |\Sigma^i|^{-\frac{n+g_i+1}{2}} \exp\left\{-\frac{1}{2}\tr\left( nS_n^i (\Sigma^i)^{-1}\right) \right\}
\end{equation}

where $S_n^i$ and $\Sigma^i$ are the sample covariance and covariance component of the $i$th clique, and $\Gamma_{g_i}(\cdot)$ is the $g_i$ dimensional multivariate gamma function.

\subsection{Jacobian for the general case}
Now assume the general case with a sparse model $\cM$ that does not have to be a clique model. Denote the $ij$th entry of $A$ as $A_{ij}$. 
Define 
\begin{equation}
S_i = \{j: A_{ij}\equiv0, j=1,\cdots,p  \},\;i=1,\cdots,p.
\end{equation}
The set $S_i$ indicates which entries of $A$ in the $i$th row are fixed at zero. 

Then equation (\ref{eq:bigJ}) becomes 
\begin{equation}\label{eq:generalbigJ}
J(\by,A)= D \left(\det\left(\tilde P\right)\right)
\end{equation}
$\tilde P = (\tilde B_1, \cdots, \tilde B_p)$ is the matrix $P$ with correct corresponding columns dropped, i.e. block $\tilde B_i$ is obtained from block $ B_i$ with $S_i$ columns removed.

Let $p_i$ be the number of nonzero entries in the $i$th row of $A$, and $U_i$ be the sub-matrix of $U$ excluding columns in $S_i$, i.e. $U_{i}=U_{[:,-S_i]}$. The exact form of Eq (\ref{eq:generalbigJ}) is the following:

\begin{itemize}
        \item when the $l_2$-norm is used,
        \begin{equation} \label{eq:L2J}
        J(\by,A) = \sqrt{\prod_i^p\det(U_i^TU_i/n)}\ ;
        \end{equation}
        
        \item if choose the $l_\infty$-norm,
        \begin{equation}\label{eq:LinftyJ}
        J(\by,A) = \left(\prod_{i=1}^p\begin{pmatrix}
        n\\p_i
        \end{pmatrix}  \right)^{-1} \Jusum \prod_{i=1}^p\left|\det\left(U_{i}\right)_{\i_i} \right|.
        \end{equation}
        
        Notice that for each $\i$, we sum over total $\prod_{i=1}^p\begin{pmatrix} n\\ p_i \end{pmatrix}$ determinant products. It can be shown by induction that Eq (\ref{eq:LinftyJ}) is equivalent to the following:
        \begin{equation}\label{eq:Jhat}
        J(\by,A)=
        \prod_{i=1}^p \overline{\left|\det\left(U_{i}\right)_{\i_i} \right|},
        \end{equation}
        where $\overline{\left|\det\left(U_{i}\right)_{\i_i} \right|}$ denotes the average absolute determinant of all possible expressions of $(U_{i})_{\i_i}$ for a row number $i$. 
\end{itemize}

\section{Consistency of fiducial distribution}\label{sec:theory}

In general, there is no one-to-one correspondence between the covariance matrix $\Sigma$ and the covariate matrix $A$. This leads to the identifiability issue from $\Sigma$ to $A$. However, if $A$ is assumed to be sparse with the sparse locations known, then the identifiability problem often vanishes. In this section we will show that, if there is one-to-one correspondence between $\Sigma$ and $A$, then the GFD of the covariate matrix achieves a fiducial Bernstein-von Mises Theorem (Theorem \ref{thm:BvM}), which provides theoretical guarantees of asymptotic normality and asymptotic efficiency for the GFD  \cite{Hannig2015}. 

The results here are derived based on FM-distance \cite{Forstner1999}. For two symmetric positive definite matrices $M$ and $N$, with the eigenvalues $\lambda_i(M,N)$ from $\det(\lambda M-N)=0$, the FM-distance between the two matrices $M$ and $N$ is 
\begin{equation}\label{eq:Forstner0}  
\d(M, N)=\sqrt{\sum_{i=1}^n\ln^2\lambda_i(M,N)}.
\end{equation}
This distance measure is a metric and invariant with respect to both affine transformations of the coordinate system and an inversion of the matrices \cite{Forstner1999}. 

The particular choice of the distance measure for covariance matrices is not crucial in proving the consistency of the GFD.

\begin{defi}
For a fixed covariate matrix $A_0$ and $\de\geq0$, define the $\de$-neighborhood of $A_0$ as the set $B(A_0,\de)=\{A: \d(AA^T, A_0A_0^T)\leq \de\}$. 
\end{defi}

Before presenting the theorem on consistency of the GFD, we will establish some regularity condition on the likelihood function and Jacobian formula (Prop \ref{prop:LA}, \ref{prop:minLA}, \ref{prop:J0}). The proofs can be found in the appendix. 

\begin{prop}{\label{prop:LA}}
For any $\de >0$ there exists $\ep>0$ such that
$$P_{A_0}\left\{ \sup_{A\not\in B(A_0,\de)}\frac{1}{n}(L_n(A)-L_n(A_0))\leq -\ep \right\}\rightarrow 1,$$
where $L_n(A) = \log f(\by,A)= \sum_{i=1}^n\log f(y_i,A).$
\end{prop}

\begin{prop}{\label{prop:minLA}}
Let $L_n(\cdot)$ be as above. Then for any $\de >0$ 
$$\inf_{A\not\in B(A_0,\de)} \frac{\min_{\substack{\i = \{i_1, \cdots, i_p\} \\ 1\leq i_1<\cdots<i_p\leq n} }\log f(A,\by_\i)}{|L_n(A)-L_n(A_0)|}\xrightarrow{A_0} 0,$$
where $ f(A,\by_\i)$ is the joint likelihood of $p$ observations $y_{i_1}, \cdots, y_{i_p}$.
\end{prop}

\begin{prop}{\label{prop:J0}}
        Assume that there is a one-to-one correspondence between $A$ and $\Sigma = AA^T$. Then the Jacobian function $J(\by, A)\xrightarrow{a. s.} \pi_{A_0}(A) $ uniformly on compacts in $A$, where $\pi_{A_0}(A)$ is a function of $A$, independent of the sample size and observations, but it depends on truth. 
\end{prop}

The Bernstein-von Mises Theorem provides conditions under which the Bayesian posterior distribution is asymptotically normal (van der Vaart 1998, Ghosh 2003). The fiducial Bernstein-von Mises Theorem is an extension that includes a list of conditions under which the GFD is asymptotically normal \cite{Sonderegger2012}. Those conditions can be divided into three parts to ensure each of the following: 
\begin{itemize}
        \item[(a)] the Maximum Likelihood Estimator (MLE) is asymptotically normal 
        \item[(b)] the Bayesian posterior distribution becomes close to that of the MLE
        \item[(c)] the fiducial distribution is close to the Bayesian posterior
\end{itemize}
 It is clear that the MLE of $f(\by,A)$ is asymptotically normal. Under our model, the conditions for (b) holds due to Proposition \ref{prop:LA} and the construction of the Jacobian formula; the conditions for (c) are satisfied by Propositions \ref{prop:minLA}, \ref{prop:J0}. Closely following \cite{Sonderegger2012}, we arrive at Theorem \ref{thm:BvM}.

\begin{thm}{\label{thm:BvM}}(Consistency) 
        Let $\mathcal{R}_A$ be an observation from the fiducial distribution $r(A|\by)$ and denote the density of $B = \sqrt{n}(\mathcal{R}_A - \hat{A}_n)$ by $\pi^*(B, \by)$, where $\hat{A}_n$ is a maximum likelihood estimator. Let $I(A)$ be the Fisher information matrix. Under the assumption that there is one-to-one correspondence from the covariance matrix $\Sigma$ to the covariate matrix $A$, 
        \begin{equation}
        \int_{\R^{p\times p}} \left | \pi^*(B, \by)- \frac{\sqrt{\det |I(A_0) |}}{\sqrt{2\pi}}\exp\{- \by^TI(A_0)\by/2\} \right |dB \xrightarrow{P_{A_0}} 0
        \label{eq:BvM0}
        \end{equation}
        
\end{thm}

Detailed proof can be found in the appendix.

\section{Model selection}\label{sec: ModelSelection}
Often time in practice, to obtain enough statistical power or simply for feasibility, sparse covariates/covariances assumptions are imposed. The exact sparse structure is usually unknown, model selection is required to determine the appropriate parameter space. 

Since GFD behaves like the likelihood function, in order to avoid over-fitting,  a penalty term $ q_\cM(n)$ on the parameter space needs to be included in the model selection process \cite{Hannig2015}. 

In this section, we will discuss how to select the parameter space under the clique model, as well as picking the correct sparse structure for the general case.

\subsection{Theoretic results for the clique models}\label{sec:cliqueresult}

Recall that under the full model, 
\begin{equation*}
r(A|\by)\propto C(\by)(2\pi)^{-\frac{np}{2}}|\det(A)|^{-(n+p)}\exp\left[-\frac{1}{2}\tr\{nS_n(AA^T)^{-1}\}  \right].
\end{equation*}

For clique model selection, we need to evaluate the normalizing constant. 
\begin{equation}\label{eq:NormalizingConstant}
\int J(\by,A)f(\by|A)\,dA
= \frac{\pi^{(p^2-np)/2}C(\by)\Gamma_p\left(\frac n2\right)}{ |\det (nS_n)|^{n/2}\Gamma_p\left(\frac p2\right)},
\end{equation}
The detailed derivation is provided in the appendix \ref{sec:normalizingconstant}.

Let us denote by $\cM$ a clique model; a collection of $k$ cliques -- sets of indexes that are related to each other. The coordinates are assumed independent if they are not in the same cliques. For any positive-definite symmetric matrix $S$, whose dimension is compatible with $\cM$, we denote $S^{\cM}$ as the matrix obtained from $S$ by setting the off-diagonal entries that corresponds to pairs of indexes not in the same clique within $\cM$ to zero. Note that $S^\cM$ is a block diagonal (after possible permutations of rows and columns) positive-definite symmetric matrix. 

The classical Fischer-Hadamard inequality \citep{Fischer1908} implies that for any positive definite symmetric matrix $S$ and any clique model
$
\det(S)\leq \det(S^\cM).
$
\cite{Ipsen2011} provides a useful lower bound: 
Let $\rho$ be the spectral radius and $\lambda$ be the smallest eigenvalue of $(S^\cM)^{-1}(S-S^\cM)$, 
\begin{equation}\label{eq:spectralbound}
e^{-\frac{p\rho^2}{1+\lambda}}\det(S^\cM)\leq \det(S)\leq \det(S^\cM).
\end{equation}

Assume the clique sizes are $g_1,\cdots, g_k.$ Then the GFD of the model is
\begin{equation}\label{eq:GFDcliqueM}
r( \cM|\by) \propto \frac{ \pi^{\frac{\sum_{i=1}^k g_i^2}2} }{|\det S_n^\cM|^{\frac n2}} \prod_{i=1}^k C_{\cM,i}(\by)\frac{\Gamma_{g_i}\left(\frac n2\right)}{\Gamma_{g_i}\left(\frac {g_i}2\right)},
\end{equation}

where $C_{\cM,i}(\by)$ denotes the Jacobian constant term (\ref{eq:CJ}) computed only using the observations in the $i$th clique. 

In the remaining part of this section we consider the dimension of $\by$ as a fixed number $p$ and the sample size $n\to\infty$. Similar arguments could be extended to $p\to\infty$ with $p/\sqrt{n}\to 0$. The proofs of the results are included in the Appendix.

\begin{lemma}\label{l:constant}
        For any clique model $\cM$ with $k$ cliques of sizes $g_i,\ i=1,\ldots k$ we have 
        \begin{itemize}
                \item[(1)] under the $l_2$ norm, $C_{\cM,i}(\by)=|\det S_{n}^{\cM, i}|^{g_i/2} $,
                $$C_{\cM,i}(\by)
                \to
                |\det (\Sigma_0^{\cM,i})|^{\frac{g_i}{2}}\; \text{a.s.};
                $$
                                                
                \item[(2)] under the $l_\infty$ norm, 
                $C_{\cM,i}(\by) = \begin{pmatrix}
                n\\g_i
                \end{pmatrix}^{-g_i}\left(\Jsumgi\left|\det\left(V_{\cM,i}\right)_{\i_i} \right|\right)^{g_i},$
                $$C_{\cM,i}(\by)
                \to 
                |\det (\Sigma_0^{\cM,i})|^{\frac{g_i}{2}} 2^{\frac{g_i^2}{2}}\pi^{-\frac{g_i}{2}}\Gamma\left(\frac{g_i + 1}{2}\right)^{g_i},\; \text{a.s.}$$ 
                where $S_{n}^{\cM, i}$ is the sample covariance computed using only observations within clique $i$ under the model $\cM$, $V_{\cM,i}$ is the sub-matrix of $V$ that only includes the observations in clique $i$, and $\Sigma_0^{\cM,i}$ denotes the $i$th block component of $ \Sigma_0^{\cM}$.
                                
        \end{itemize}   
\end{lemma}

Lemma \ref{l:constant} provides the limits of the constant $C_{\cM,i}(\by)$ as sample size increases. The next lemma shows how the ratio $\frac{\prod_{i=1}^k \Gamma_{g_i}\left(\frac{n}{2}\right)}{\prod_{j=1}^l \Gamma_{h_j}\left(\frac{n}{2}\right)}$ behaves when sample size increases. 

\begin{lemma}\label{l:gamma}
        Let $g_i, i=1,\ldots, k$ and $h_j, j=1,\ldots,l$ be integers such that $\sum_{i=1}^k g_i=\sum_{j=1}^l h_i$. Then
        as $n\to\infty$
        \[
        \frac{\prod_{i=1}^k \Gamma_{g_i}\left(\frac{n}{2}\right)}{\prod_{j=1}^l \Gamma_{h_j}\left(\frac{n}{2}\right)}
        \sim
        \left(\frac{\pi}n\right)^{\frac{\sum_{i=1}^k g_i^2-\sum_{j=1}^l h_j^2}4}.
        \]
\end{lemma}

Given two clique models $\cM_1$ and $\cM_2$. We write $\cM_1\subset \cM_2$ if cliques in $\cM_2$ are obtained by merging cliques in $\cM_1$. Consequently, $\cM_2$ has fewer cliques and these cliques are larger than $\cM_1$. 

Let $\cM_0, \Sigma_0$ be the ``true'' clique model and covariance matrix used to generate the observed data. 

We will call all the clique models $\cM$ satisfying $\Sigma_0^{\cM}=\Sigma_0$ compatible with the true covariance matrix. We assume that $\cM_0\subset \cM$ for all clique models compatible  with $\Sigma_0$. 

\begin{lemma}\label{l:detSnratio}
                Let $\cM$ be a clique model. 
                \begin{itemize}
                        \item[i.] If $\det(\Sigma_0)<\det(\Sigma_0^\cM)$, then there is $a>0$, such that
                        \[
                        \left|\frac{\det (S_n^{\cM_0})}{\det (S_n^\cM)}\right|^{n/2} \leq e^{-an}\quad\mbox{eventually a.s.}
                        \]
                        
                        \item[ii.] If $\cM\neq\cM_0$ is compatible with $\Si_0$, then as $n\to\infty$
                        \[
                        \left|\frac{\det (S_n^{\cM_0})}{\det (S_n^\cM)}\right|^{n/2} = \mathcal{O}_P(1).
                        \]
                \end{itemize}
        \end{lemma}

We are now ready to state our main theorem, which provides some guideline for choosing penalty function $q_\cM(n)$. Define the penalized GFD of the model as $r_p(\M|\by) = r(\M|\by)q_\M(n)$.
\begin{thm}\label{t:clique}
        For any clique model $\cM$ that is not compatible with $\Sigma_0$, assume $\det\left(\Sigma_0\right)<\det\left(\Sigma_0^\cM\right)$ and  the penalty  $e^{-a n} q_\cM(n)/q_{\cM_0}(n)\to 0$ for all $a>0$ as $n\to 0$.
        
        For any clique model $\cM$ compatible with $\Sigma_0$  assume that 
        $q_\cM(n)/q_{\cM_0}(n)$ is bounded.
        
        Then as $n\to\infty$ with $p$ held fixed
        $
        r_p(\cM_0|\bY)\toP 1. 
        $
\end{thm}

The exact form of the penalty function depends on the norm choice for the Jacobian. Under the $l_2$-norm, the following penalty function works the best.

\begin{equation}\label{eq:cliquepen}
q_\M(n) = \exp\left\{-\sum_{i = 1}^k \left[ \frac{1}{4}g_i^2\log(n) -  \frac{1}{2}g_i^2\log(g_i) \right]\right \}.
\end{equation} 

It is easy to check that Eq \ref{eq:cliquepen} satisfies Theorem \ref{t:clique}.

\subsection{Penalty term for the general case}
For the general case we propose the following penalty function that is based on the Minimum Description Length (MDL) \cite{Rissanen1978} for a model $\cM$:
\begin{equation}\label{eq:MDL}
q_\M(n)=\exp\left\{-\sum_{i=1}^p\left[\frac{1}{2}p_i\log(np) + \log \begin{pmatrix} p\\p_i\end{pmatrix} \right]\right \}
\end{equation}
where $\M$ corresponds to a $p\times p$ matrix with $p_i$ many non-fixed-zero elements in its $i$th row, and $n$ is the number of observations. 

The penalized GFD of $A$ is therefore, 
\begin{equation}\label{eq:pGFDgeneral}
r_p(A|\M, \by) \propto r(A|\M,\by)\exp\left\{-\sum_{i=1}^p\left[\frac{1}{2}p_i\log(np) + \log \begin{pmatrix} p\\p_i\end{pmatrix} \right]\right \}.
\end{equation}

\section{Sample from the GFD}\label{sec:simulation} 

 Given the true model $\cM_0$, standard Markov chain Monte Carlo (MCMC) methods can be utilized for the estimation of the covariance matrix. Under the full model and clique model, the GFD of $\Sigma$ follows either a inverse Wishart distribution or a composite of inverse Wishart distributions (see Section \ref{sec:COVmodel}). Sampling from the GFD becomes straight forward and it can be done through one of the inverse Wishart random generation functions, e.g. InvWishart (MCMCpack, R) or iwishrnd (Matlab). 
 
 When $p$ is small and $n$ is large, the estimation of $\Sigma$ can always be done through this setting, regardless if there are zero entries in $A$. The concept of having entries of $A$ fixed at zero is to impose sparsity structure and allow estimation under a high dimensional setting without requiring large number of observations. As in practice the true sparse structure is often unobserved, we will focus on the cases where $\cM_0$ is not given.

 \subsection{Sample from a clique model}
 Estimation of cliques is closely related to applications in network analysis, such as communities of people in social networks and gene regulatory network. Recall the penalized clique model GFD introduced in Section \ref{sec:cliqueresult},
\begin{equation*}
r_p( \cM|\by) \propto \frac{ \pi^{\frac{\sum_{i=1}^k g_i^2}2} }{|\det S_n^\cM|^{\frac n2}} \prod_{i=1}^k C_{\cM,i}(\by)\frac{\Gamma_{g_i}\left(\frac n2\right)}{\Gamma_{g_i}\left(\frac {g_i}2\right)}q_\cM(n),
\end{equation*}

\begin{SCfigure}[][h]
                \includegraphics[height = 9cm]{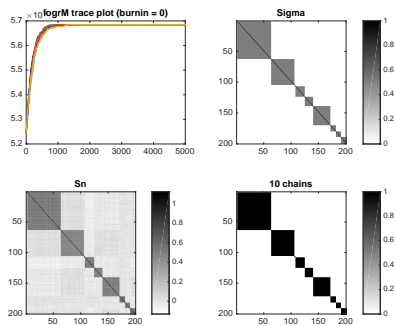}
                \vspace{-.9in}
                \caption{\label{fig:Cliquep200n1000} Result for $k =10, p = 200, n = 1000$. The trace plot (top left) shows that the chains converge quickly. Although$\frac{n}{p}$ is small, the sample covariance (bottom left) roughly captures the shape of true covariance (top right). The last panel (bottom right) shows that the fiducial estimate captures the true clique structure perfectly. }
        \end{SCfigure}

 Assuming that both the number of cliques $k$ and the clique sizes $g_k$'s are unknown, the clique structure can be estimated via Gibbs sampler. Example 1 shows the simulation result for a $200 \times 200$ covariance matrix. We consider the covariance matrix to be with 1's on the diagonal and $ij$th entry being 0.5 if the coordinate $i$ and $j$ belongs to a clique. From top down, left to right, Figure \ref{fig:Cliquep200n1000} shows the trace plot for $\log(r_p( \cM|\by))$ without normalizing constant, true covariance $\Sigma$, sample covariance $S_n$, and the fiducial probability of the estimated cliques based on the 10 Gibbs sampler Markov chains with random initial states. The trace plot helps to monitor the convergence. The fiducial probability of cliques panel reveals the clique structure precisely. The last panel is the aggregate result of 4000 iterations with burn in = 1000 from the 10 Markov chains.

\begin{SCfigure}[][h]
                \includegraphics[height = 9cm]{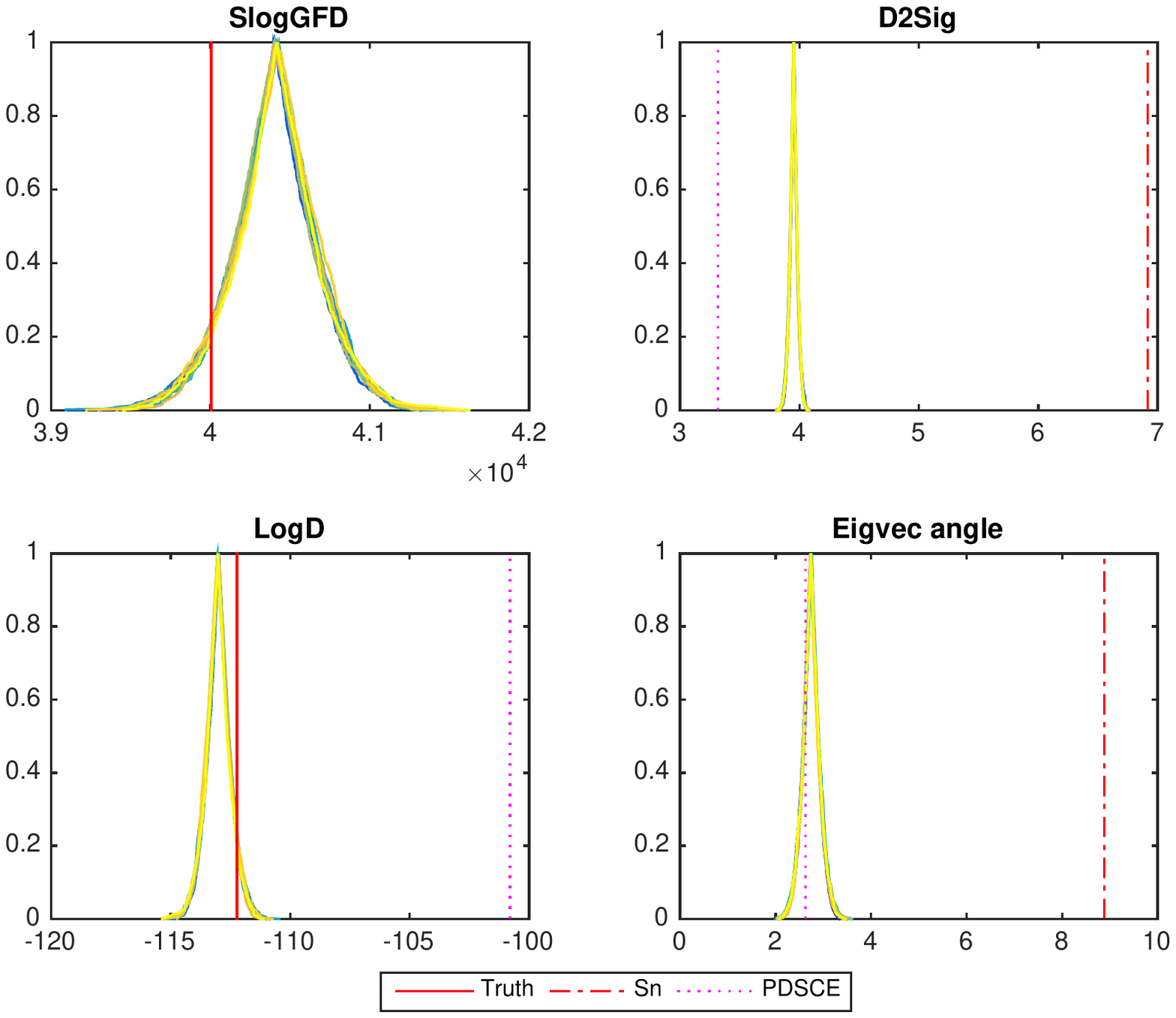}
                \vspace{-.9in}
                \caption{\label{fig:Cliquep200n1000cc} Confidence curve plots for estimated covariance matrix. $k =10, p = 200, n = 1000$. Comparing to the sample covariance, the estimators are closer to $\Sigma$. The PDSCE estimator shows even smaller FM-distance to $\Si$, it, however, greatly overestimates $\det{\Si}$. }
        \end{SCfigure}

 The covariance estimators can be obtained by sampling from inverse Wishart distributions based on the estimated clique structure. Figure \ref{fig:Cliquep200n1000cc} shows the confidence curves of four statistics for estimated covariance matrix $\hat{\Sigma}$: log-transformed generalized fiducial likelihood (SlogGFD), distance to $\Sigma$ (D2Sig), log-determinant (LogD), and angle between the leading eigenvectors of $\hat{\Sigma}$ and $\Sigma$ (Eigvec angle). The truth for SlogGFD and LogD are shown as red solid vertical lines. In D2Sig and Eigvec angle panels, we include comparisons to sample covariance as red dotted-dashed vertical lines. In addition, we compute the point estimation via the Positive Definite Sparse Covariance Estimators (PDSCE) method introduced in \cite{Rothman2012}. Its corresponding statistics are shown as magenta dotted vertical lines. In this example, the fiducial estimates peak near the truth in Panels SlogGFD and LogD. The estimated covariance matrices all appear to be more similar to $\Sigma$ than $S_n$ as shown in panels D2Sig and Eigvec angle. The PDSCE estimator is even closer to $\Sigma $ in terms of FM-distance; it however greatly overestimates $\det{\Si}$.
  
 The PDSCE method produces a good point estimator to the covariance matrix. It is worth noting that our method shows similar performance with the benefit of producing a distribution of estimators.
 
\begin{SCfigure}[][h]
                \includegraphics[height = 9cm]{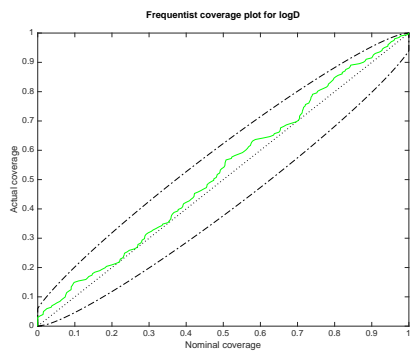}
                \vspace{-.9in}
                \caption{\label{fig:Cliquep200n1000logD} 95\% coverage plots for 100 repeated simulations. $k =10, p = 200, n = 1000$. The p-values (in green) roughly follow a uniform [0,1] distribution, and they lie inside of the envelope.  }
        \end{SCfigure}

 We repeat the clique simulation 200 times, each with one Markov chain started at random, and compute the one-sided p-values for the estimate covariance log determinant. With the same true covariance matrix, a new set of 1000 observations are generated for each simulation. Figure \ref{fig:Cliquep200n1000logD} shows the quantile-quantile plot of the p-values against the uniform [0,1] distribution. The dotted-dashed envelope is the 95\% coverage band. It shows a well-calibrated 95\% confidence interval. The p-value curve (in green) is well enclosed by the envelope, indicating good calibration of the  coverage. 

 \subsection{Sampling in the general case}\label{Sec:SimMCMC}

  For the general case, if the sparse model is unknown, we propose to utilize a reversible jump MCMC (RJMCMC) method to efficiently sample from Eq (\ref{eq:pGFDgeneral}) and simultaneously update $\M$.
  
  RJMCMC is an extension of standard Markov chain Monte Carlo methods that allows simulation of the target distribution on spaces of varying dimensions \cite{Green1995}. The ``jumps'' refers to moves between models with possibly different parameter spaces. More details on RJMCMC can be found in \cite{shi2015}. Since $\cM$ is unknown, namely the number and the locations of fixed zeros in the matrix $A$ is unknown, the property of jumping between parameter spaces with different dimension is desired for estimating $\Sigma = A A^T$. Because the search space for RJMCMC is both within parameter space and between spaces, it is known for slower convergence. To improve efficiency of the algorithm, we adapt the zeroth-order method \cite{Brooks2003} and impose additional sparse constrains.

 Assuming that there are fixed zeros in $A$, then for a $p\times p$ matrix $A$, the number needed to be estimated is less than $p^2$. If there are many fixed zeros, then this number is much smaller, hence the estimation is feasible even if the number of observations $n$ is less than $p$. In other words, the sparsity assumption on $A$ allows estimations under a large $p$ small $n$ setting. Suppose the zero entry locations of $A$ are known. The rest of $A$ can be obtain via standard MCMC techniques, such as Metroplis-Hastings.

\begin{SCfigure}[][h]
        \includegraphics[height = 9cm]{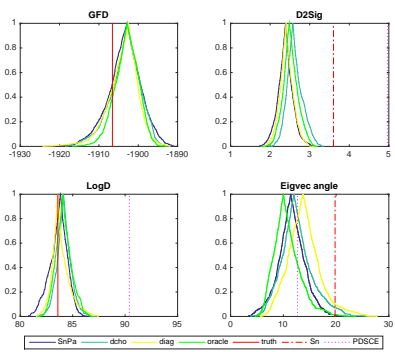}
        \vspace{-.9in}
        \caption{\label{fig:MCMCp15n30} All the chains show good estimation of covariance matrix. The estimators are better than both the sample covariance matrix and the PDSCE estimator. }
 \end{SCfigure}

 Figure \ref{fig:MCMCp15n30} considers a case with $p = 15,\; n = 30$. It shows the confidence curve plot per Markov chain for each statistic of interest. In addition to D2Sig, LogD, and Eigvec angle as before, we have GFD ($\log(r_p(A|\by))$ without the normalizing constant). The initial states for the four Markov chains are SnPa ($S_n$ restricted to $maxC$ (see Section {\ref{Sec:SimRJMCMC}}), in blue), dcho(diagonal matrix of Cholesky decomposition, in cyan), diag (diagonal matrix of $S_n$, in yellow), and oracle (true $A$, in green). In addition, we include the statistics for $\Si$, $S_n$, and the PDSCE estimator in comparison with the confidence curves. They are shown as vertical lines as in the previous example.
 
The fiducial estimators have confidence curves peak around the truth in Panels GFD and LogD. In the right two panels, the (majority of) fiducial estimators lie on the left of the dotted-dashed lines, indicating that the estimators are closer to the truth than the sample covariance. The PDSCE estimator falls on the right edge of the Panel D2Sig shows that it is not as close to the truth. As before, the PDSCE estimator overestimates the covariance determinant. Here, burn in = 5000, window = 10000.

 \subsection{General case with sparse locations unknown}\label{Sec:SimRJMCMC}
In the general case with sparse locations unknown, we further assume that there is a maximum number of nonzeros per column allowed, denoted as $maxC$. This additional constraint can be viewed as each predictor only contribute to few tuples of the multivariate response. This assumption has been implemented to reduce the search space for RJMCMC. The starting states include MaxC ($S_n^{0.5}$, restricted to $maxC$, in blue) along with chol (in cyan), dcho (in artichoke), diag (in yellow), and true (in green) as before. We will revisit the example discussed in Section \ref{Sec:SimMCMC}. 

\begin{SCfigure}[][h]
        \includegraphics[height = 9cm]{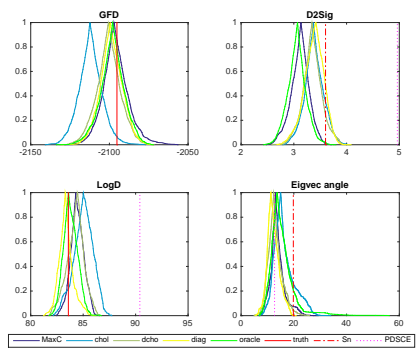}
        \vspace{-.9in}
        \caption{\label{fig:RJp15n30} Similar to Figure \ref{fig:MCMCp15n30}, the fiducial estimators are better than both the sample covariance matrix and the PDSCE estimator in this case. }
\end{SCfigure}

(See Figure \ref{fig:RJp15n30}). In the left two panels, the fiducial estimators peak at the true fiducial likelihood and covariance determinant. The distance comparison plot (top right) show that the estimators are closer to the truth than both the sample covariance matrix and the PDSCE estimator. Bottom right panel shows that the leading eigenvector of the estimators are as close to the truth as for sample covariance and the PDSCE estimator as in Figure \ref{fig:MCMCp15n30}. Here, burn in = 50000, window = 10000. \\

Additional simulations are included in appendix. All simulations shown here use the $l_2$-norm. The implementation for the $l_\infty$-norm is similar, but computationally more expensive.

\section{Discussion}\label{sec:COVdiscussion}

Covariance estimation is an important problem in statistics. There have been much effort made toward it both in the fields of Bayesian and frequentist. In this manuscript we propose to look into this classical problem via a generalized fiducial approach. We demonstrate that, under mild assumptions, the generalized fiducial distribution of the covariate matrix is asymptotically normal. In addition, we discuss the clique model and show that the fiducial approach is a powerful tool for identifying clique structures, even when the dimension of the parameter space is large and the ratio $n/p$ is small. To identify the covariance structure for non-clique models, in contrast to typical sparse covariance/precision matrix assumptions, we look at cases where the ratio $n/p$ is small and the covariate matrix is sparse. This ``unusual'' sparsity assumption arises in applications where multiple dependent variables contribute to several response variables collaboratively. The fiducial approach allows us to obtain a distribution of covariance estimators that are better than sample covariance, and comparable to the PDSCE estimator. The distances to true covariance matrix show that as dimension increases, the fiducial estimators become closer to the true covariance matrix.

Similar to Bayesian approaches, generalized fiducial inference produces a distribution of estimators, yet the two methods differ fundamentally. Bayesian methods rely on prior distributions on the parameter of interest, while fiducial approaches depend on the data generating equation. In the framework discussed here, the data generating mechanism is natural to establish than choosing appropriate priors while some other times priors are easier to construct. 
  
Estimating sparse covariance matrix without knowing the fixed zeros is a hard problem. While our approach shows promising results for the clique model, for the general case it still suffers from a few drawbacks: (1) due to the nature of RJMCMC, the computational burden can be significant if the matrix is not very sparse; (2) to limit the search space, a row/column-wise sparsity upper bound needs to be chosen based on prior knowledge of the data type; (3) the results presented in this manuscript assume a squared covariate matrix, which can be limited to direct applications to high-throughput data. Furthermore, a more sophisticated way of choosing initial states and mixing method can improve the efficiency of our algorithm. It is possible and well-worth it to extend our current work to more general cases. 

\section{Appendix}\label{sec:App}

\subsection{Proof of Proposition \ref{prop:LA}}

\begin{prop*}[\ref{prop:LA}]
        For any $\de >0$ there exists $\ep>0$ such that
        $$P_{A_0}\left\{ \sup_{A\not\in B(A_0,\de)}\frac{1}{n}(L_n(A)-L_n(A_0))\leq -\ep \right\}\rightarrow 1,$$
        where $L_n(A) = \log f(\by,A)= \sum_{i=1}^n\log f(Y_i,A).$
\end{prop*}
\begin{proof}
        Let $\Si = AA^T,\;\Si_0 = A_0A_0^T$. Denote $S_n$ as the sample covariance matrix as before, $n\in \mathbb{N}$. Since $S_n$ is the maximum likelihood estimator, we have 
        $$S_n\xrightarrow{P_{A_0}} \Si_0, $$
        $$\text{i.e. } \forall r>0,\; P_{A_0}(\{\om:\d(S_n(\om),\Si_0)\geq r \})\rightarrow 0.$$ 
        Define $L_{\de,n} = \{\om:\d(S_n(\om),\Si_0)< \de/2 \}$. For an arbitrary $\om \in L_{\de,n}, $ assume that $\la_i^\dagger$'s and $\la_i^\ast$'s are the eigenvalues of $S_n(\om)\Si^{-1}$ and $S_n(\om)\Si_0^{-1}$, respectively. Suppose that $A\not\in B(A_0,\de)$, then
        $$\de<\d(\Si,\Si_0)\leq \d(\Si,S_n(\om)) + \d(S_n(\om),\Si_0)< \d(\Si,S_n(\om)) + \de/2$$
        $$\Rightarrow \d(\Si,S_n(\om)) =\sqrt{\sum_{i=1}^p \ln^2\la_i^\dagger}> \de/2$$
        So there exists $k\in\{1,2,\cdots,p\}$, such that $\ln^2\la_k^\dagger>\frac{\de^2}{4p}$, then 
        $$\ln\la_k - \la_k<\max\left\{ \frac{\de}{2\sqrt{p}}-\exp\left(\frac{\de}{2\sqrt{p}}\right) , -\frac{\de}{2\sqrt{p}}-\exp\left(-\frac{\de}{2\sqrt{p}}\right) \right\}:=m_\de,$$
        due to the fact that the function $g(\la) = \ln\la - \la$ is concave with unique maxima $\la = 1$; $g(1)=-1$. \\
        Meanwhile, 
        \begin{eqnarray*}
                &&\frac{1}{n}(L_n(A)-L_n(A_0))(\om)\\
                &=&-\ln|\det(A)|-\half\tr\{Sn(\om)\Si^{-1}\}+\ln|\det(A_0)|+\half\tr\{Sn(\om)\Si_0^{-1}\}\\
                &=&\half \ln(Sn(\om)\Si^{-1})-\half\tr\{Sn(\om)\Si^{-1}\}-\half \ln(Sn(\om)\Si_0^{-1})+\half\tr\{Sn(\om)\Si_0^{-1}\}\\
                &=&\half\left\{\sum_{i=1}^p(\ln\la_i^\dagger-\la_i^\dagger) - \sum_{i=1}^p(\ln\la_i^\ast-\la_i^\ast) \right\}\\
                &<& \half\left\{-(p-1)+m_\de+p\right\}\\
                &=& \half(m_\de+1)
        \end{eqnarray*}
        $$\Rightarrow \sup_{A\not\in B(A_0,\de)}\frac{1}{n}(L_n(A)-L_n(A_0))(\om)\leq  \half(m_\de+1)<0.$$
        Let $\ep =- \half(m_\de+1),\; U_{\de,n}=\left\{ \om:\sup_{A\not\in B(A_0,\de)}\frac{1}{n}(L_n(A)-L_n(A_0))(\om)\leq -\ep \right\}$. Then $L_{\de,n}\subseteq U_{\de,n}$. Notice that 
        $$1=\lim_{n\rightarrow \infty} P_{A_0}(L_{\de,n})=\liminf_{n\rightarrow \infty} P_{A_0}(L_{\de,n})\leq\liminf_{n\rightarrow \infty} P_{A_0}(U_{\de,n})\leq\limsup_{n\rightarrow \infty} P_{A_0}(U_{\de,n})\leq1,$$
        Therefore, $\lim_{n\rightarrow \infty} P_{A_0}(U_{\de,n})=1.$  
\end{proof}

\subsection{Proof of Proposition \ref{prop:minLA}}

\begin{prop*}[\ref{prop:minLA}]
        Let $L_n(\cdot)$ be as above. Then for any $\de >0$ 
        $$\inf_{A\not\in B(A_0,\de)} \frac{\min_{\substack{\i = \{i_1, \cdots, i_p\} \\ 1\leq i_1<\cdots<i_p\leq n} }\log f(A,\by_\i)}{|L_n(A)-L_n(A_0)|}\xrightarrow{A_0} 0,$$
        where $ f(A,\by_\i)$ is the joint likelihood of $p$ observations $y_{i_1}, \cdots, y_{i_p}$.
\end{prop*}

\begin{proof}
        Note that 
        $$\inf_{A\not\in B(A_0,\de)} \frac{\min_{\substack{\i = \{i_1, \cdots, i_p\} \\ 1\leq i_1<\cdots<i_p\leq n} }\log f(A,\by_\i)}{|L_n(A)-L_n(A_0)|} \leq \frac{ \inf_{A\not\in B(A_0,\de)} \min_{\substack{\i = \{i_1, \cdots, i_p\} \\ 1\leq i_1<\cdots<i_p\leq n} }\log f(A,\by_\i)}{ \inf_{A\not\in B(A_0,\de)} |L_n(A)-L_n(A_0)|}$$
        For any $A\not\in B(A_0,\de)$, denote $\Si = AA^T,\;\Si_0 = A_0A_0^T$ and let $t>0$, we have  
        \begin{eqnarray*}
                &&P_{A_0}\left(\min_{\substack{\i = \{i_1, \cdots, i_p\} \\ 1\leq i_1<\cdots<i_p\leq n} }\log f(A,\by_\i) \leq -t \log n \right)\\
                &\leq&P_{A_0}\left(\min_{i=1,\cdots, n}\log f(A,Y_i) \leq -\frac{t\log n}{p}  \right)\\
                &=& 1 - \left[1 - P_{A_0}\left( - \log f(A,Y_i) \geq -\frac{t\log n}{p}  \right)\right]^n\\
                &\leq& 1 - \left[1 - \frac{p\E_{A_0} ( - \log f(A,Y_i)  )}{t\log n}\right]^n\;\;\;\;\;\text{ (Markov inequality)}\\
                &=& 1 - \left[1 - \frac{p(\log(2\pi) + \log\det(\Si) + \tr\{\Si^{-1}\Si_0 \}) }{2t\log n}\right]^n\\
                &\rightarrow& 0, \;\;\;\text{  as }n\rightarrow \infty.
        \end{eqnarray*}
        Note that the numerator goes to $-\infty$ at most as fast as $-t\log n$. Meanwhile, for a fixed $n$ and any $\om \in L_{\de,n} = \{\om:\d(S_n(\om),\Si_0)< \de/2 \}$, 
        $$\inf_{A\not\in B(A_0,\de)} |L_n(A)-L_n(A_0)| = -\sup_{A\not\in B(A_0,\de)}L_n(A)-L_n(A_0) \geq \ep n$$
        By Proposition (\ref{prop:LA}), 
        $$\lim_{n\rightarrow \infty}P_{A_0}\left(\inf_{A\not\in B(A_0,\de)}|L_n(A)-L_n(A_0)|   \geq \ep n\right)= 1, $$
        i.e. the denominator goes to infinity at least as fast as $\ep n$. 
\end{proof}

\subsection{Proof of Proposition \ref{prop:J0}}

\begin{prop*}[\ref{prop:J0}]
        Assume that there is a one-to-one correspondence between $A$ and $\Sigma = AA^T$. Then the Jacobian function $J(\by, A)\xrightarrow{a. s.} \pi_{A_0}(A) $ uniformly on compacts in $A$, where $\pi_{A_0}(A)$ is a function of $A$, independent of the sample size and observations. 
\end{prop*}

\begin{proof}
        This proposition states that the Jacobian function is a U-statistic. We will show the proofs for the $l_2$-norm and the $l_\infty$-norm. For a different norm choice, the proof varies slightly.\\
        Given an ordered index vector $\r=(r_1,\cdots,r_l)$, let $E_\r=(e_{r_1}; \cdots; e_{r_l})$, where each $e_{r_j}$ is a $1\times p$ vector with 1 in the $r_j$th tuple and 0 everywhere else. Denote $-{\bf r} = \{1, \cdots, p\}\setminus {\bf r}$.\\
        Under the $l_2$-norm, 
        \begin{equation*}
        J(\by, A) = \sqrt{\prod_{i=1}^p \det(U_i^TU_i/n) }=\sqrt{\prod_{i=1}^p \det\left(E_{-S_i}^TA^{-1}S_n(A^{-1})^TE_{-S_i}\right) }
        \end{equation*}
        where $S_i$ is the list of indexes of fixed zeros in the $i$tn row of $A$.
        
        By the Strong Law of Large Numbers and follows by continuity, 
        $$J(\by, A) \longrightarrow  \sqrt{\prod_{i=1}^p \det\left(E_{-S_i}^TA^{-1}\Sigma_0(A^{-1})^TE_{-S_i}\right) } := \pi_{A_0}(A)\;\;\text{ a.s.}$$
        
        Note that both $P_n= J(\by, A)$ and $P_0=\pi_{A_0}(A)$ are polynomials of entries of $A^{-1}.$ If $A$ is in compact, the coefficients of $P_n$ converge to the coefficients of $P_0$ uniformly. Furthermore, the derivative is bounded, hence $P_n$ is equicontinuous. We have $J(\by, A)\xrightarrow{a. s.} \pi_{A_0}(A) $ uniformly on compacts in $A$.

        Under the $l_\infty$-norm, let $\bY_0 = (Y_1, Y_2, \cdots, Y_p)$ and $\pi_{A_0}(A) = \E_{A_0}J(\bY_0,A)$.
        $$J(\by,A)=
        \prod_{i=1}^p \overline{\left|\det\left(U_{i}\right)_{\i_i} \right|}$$
                
        By Theorem 1 of Yeo and Johnson \cite{Yeo2001}, it suffices to show the following:\\
        For $j = 1,\cdots, p$, set 
        $$J_j((y_1,\cdots,y_j),A) = \E_{A_0}J((y_1,\cdots,y_j, Y_{j+1},\cdots, Y_p),A).$$
        \begin{itemize}
                \item[(\ref{prop:J0}a)] There is an integrable and symmetric kernel $g(\cdot)$ and compact space $\bar B(A_0,\de)$ such that, for all $A$, and ${\bf y}_0 =(y_1,\cdots,y_p) \in\mathbb{R}^{p\times p}$,
                $$|J(\by_0,A)| \leq g(\by_0).$$
                
                \item[(\ref{prop:J0}b)] There is a sequence $S_M^p$ of measurable sets such that
                $$P\left(\mathbb{R}^{p\times p} - \bigcup_{M=1}^\infty S_M^p\right) = 0$$
                
                \item[(\ref{prop:J0}c)] For each $M$ and for all $j = 1,\cdots,p,\;J_j((y_1,\cdots,y_j),A)$ is equicontinuous in $A$ for $(y_1,\cdots,y_j)\in S^j_M$, where $S^p_M = S^j_M\times S^{p-j}_M$. 
                
        \end{itemize}
        Denote 
        $$W_{j} = \begin{pmatrix}(A^{-1}_0y_1)^T\\ \vdots\\ (A^{-1}_0y_j)^T\\ (A^{-1}_0Y_{j+1})^T \\ \vdots\\ (A^{-1}_0Y_p)^T \end{pmatrix}  = \begin{pmatrix}z_1^T\\\vdots \\ z_j^T\\Z_{j+1}^T\\\vdots \\ Z_p^T \end{pmatrix}.$$
        Then 
        $$U_0 =  \begin{pmatrix}
        (A^{-1}y_{1})^T \\ \vdots\\ (A^{-1}y_{p})^T 
        \end{pmatrix}
        = W_0 (A^{-1}A_0)^T.
        $$
        For the remaining part of the proof, let $\sum_{\r_i}$ and $\sum_{\i}$ be short for $\sum_{\substack{ \r_i=(i_{i,1},\cdots,i_{i,p_i}),\\1\leq i_{i,1}<\cdots<i_{i,p_i}\leq p}}$ and  $\sum_{\substack{\i=(\i_1, \cdots, \i_p)\\ \forall i,\; \i_i=(i_{i,1},\cdots,i_{i,p_i}),\\1\leq i_{i,1}<\cdots<i_{i,p_i}\leq p}}$, respectively.  
        
         $$J(\by_0,A) = \prod_{i=1}^p \overline{\left|\det\left(U_{0,i}\right)_{\i_i} \right|}
        \leq \prod_{i=1}^p \max\left\{\left|\det\left(U_{0,i}\right)_{\i_i} \right|\right\}.$$ Without loss of generality, let $\i_i$ denote the index set that maximizes $\left|\det\left(U_{0,i}\right)_{\i_i} \right|$.
        
        Using the Cauchy-Binet formula, 
        \begin{eqnarray*}
        \det\left(U_{0,i}\right)_{\i_i}  &=& \det\left([ W_0 (A^{-1}A_0)^T]_{(\i_i,-S_i)}\right)\\
                &=&  \sum_{\r_i} \det\left( [W_0]_{(\i_i,\r_i)}\right) \det\left( [A^{-1}A_0]_{(-S_i,\r_i)}\right).
        \end{eqnarray*}
        
        \begin{eqnarray*}
        \Rightarrow     J(\by_0,A) &\leq&   \prod_{i=1}^p\left|  \sum_{\r_i} \det\left( [W_0]_{(\i_i,\r_i)}\right) \det\left( [A^{-1}A_0]_{(-S_i,\r_i)}\right)\right|\\
                &\leq&  \prod_{i=1}^p\left\{ \max_{\r_i'}  \left|\det\left( [W_0]_{(\i_i,\r_i')}\right)  \right| \sum_{\r_i}\left|\det\left( [A^{-1}A_0]_{(-S_i,\r_i)}\right)\right|\right\}
                        \end{eqnarray*}

        Define $I_\r= E_\r E_\r^T$ to be the matrix similar to an identity matrix but with the $kk$th entry being 0 if $k\not\in\r= \{r_1,\cdots,r_l\}$. Then 
        $$\left|\det\left( [A^{-1}A_0]_{(-S_i,\r_i)}\right) \right|= \sqrt{\det\left[E_{-S_i}^T(A^{-1}A_0)I_{\r_i}(A^{-1}A_0)^TE_{-S_i} \right]}$$
        
        By Hadamard's inequality, 
        \begin{eqnarray*}
                && \det\left[E_{-S_i}^T(A^{-1}A_0)I_{\r_i}(A^{-1}A_0)^TE_{-S_i}\right]\\
                &\leq& \prod_{k\notin S_i} \left[ (A^{-1}A_0)I_{\r_i}(A^{-1}A_0)^T \right]_{kk}\\
                &\leq& \prod_{k\notin S_i} \left[ (A^{-1}A_0)(A^{-1}A_0)^T \right]_{kk}
        \end{eqnarray*}
        
        Furthermore, recall that $A\in \bar B(A_0, \delta)$, i.e. $\d(A,A_0)\leq\delta$. Then $\forall k$,
        \begin{eqnarray*}
                \left[ (A^{-1}A_0)(A^{-1}A_0)^T \right]_{kk} &=& e_k^T(A^{-1}A_0)(A^{-1}A_0)^Te_k\\
                &=& e_k^TA^{T}(\Sigma^{-1}\Sigma_0)(A^{-1})^Te_k\\
                &\leq& \la_1
                \leq e^{\sqrt{\ln^2\la_1}}
                \leq e^\delta, 
        \end{eqnarray*}
        where $\la_1$ is the leading eigenvalue of $\Sigma^{-1}\Sigma_0$.

        Hence, $$J(\by_0,A) \leq \prod_{i=1}^p\left\{\begin{pmatrix}p\\p_i\end{pmatrix} e^{\delta p_i/2}\max_{\r_i}\left\{  \left|\det\left( [W_0]_{(\i_i,\r_i)}\right)  \right| \right \}\right\} := g(\by_0)$$
        
        It is clear that $g(\by_0)$ is integrable and symmetric. \\
        
        Note that 
        \begin{eqnarray*}
                J_j((y_1,\cdots,y_j), A) &=& \E_{A_0}{J((y_1,\cdots,y_j, Y_{j+1}, \cdots, Y_p), A)}\\
                &=&\E\left\{  \prod_{i=1}^p\left\{\sum_{\r_i}\left|\det\left([ W_j (A^{-1}A_0)^T]_{(\r_i,-R_i)}\right) \right| \right\}\right\}\\
                &=&\E\left\{  \prod_{i=1}^p\left\{\sum_{\r_i}\left|\det\left( E_{\r_i}W_j (A^{-1}A_0)^T E_{-R_i}\right) \right| \right\}\right\}
        \end{eqnarray*}
        can be viewed as a polynomial function of entries of $A^{-1}$ with coefficients being polynomial functions of $y_1,\cdots,y_j$. \\
        Let $S_M^p = \{(y_1,\cdots, y_p): |y_i|\leq M, \forall i\}$, where $M$ is a positive integer. It is clear that $S_M^p$'s are measurable. By construction, $P\left(\mathbb{R}^{p\times p} - \bigcup_{M=1}^\infty S_M^p\right) = 0$. For each fixed $M$, if $(y_1,\cdots,y_j)\in S_M^j $, then the coefficients of $J_j((y_1,\cdots,y_j), A)$ are bounded, hence $J_j((y_1,\cdots,y_j), A)$ is equicontinuous in $A$.
        
\end{proof}

\subsection{Proof of Theorem \ref{thm:BvM}}    
\begin{thm*}[\ref{thm:BvM}]
        Let $\mathcal{R}_A$ be an observation from the fiducial distribution $r(A|\by)$ and denote the density of $B = \sqrt{n}(\mathcal{R}_A - \hat{A}_n)$ by $\pi^*(B, \by)$, where $\hat{A}_n$ is a maximum likelihood estimator based on $\by$. Let $I(A)$ be the Fisher information matrix. Under the assumption that there is one-to-one correspondence from the covariance matrix $\Sigma$ to the covariate matrix $A$, 
        $$\int_{\R^{p\times p}} \left | \pi^*(B, \by)- \frac{\sqrt{\det |I(A_0) |}}{\sqrt{2\pi}}\exp\{- \by^TI(A_0)\by/2\} \right |dB \xrightarrow{P_{A_0}} 0$$
\end{thm*}

\begin{proof}
        Proposition \ref{prop:J0} and the uniform strong law of large numbers for U-statistics imply that $\pi_{A_0}(A)$ is continuous, 
        $$\sup_{A\in B(A_0,\delta)} |J(\by,A) - \pi_{A_0}(A)| \rightarrow 0 \;\;\;a.s.\;P_{A_0}$$
        \begin{align*}
        \pi^*(B, \by) & = \frac{J\left(\by, \hat{A}_n + \frac{B}{\sqrt{n}}\right)f\left(\by|\hat{A}_n + \frac{B}{\sqrt{n}}\right)}{\int_{\R^{p\times p}}J\left(\by, \hat{A}_n + \frac{C}{\sqrt{n}}\right)f\left(\by|\hat{A}_n + \frac{C}{\sqrt{n}}\right)dC}\\
        & = \frac{J\left(\by, \hat{A}_n + \frac{B}{\sqrt{n}}\right) \exp\left[L_n\left(\hat{A}_n + \frac{B}{\sqrt{n}}\right) - L_n(\hat{A_n})\right]}{\int_{\R^{p\times p}}J\left(\by, \hat{A}_n + \frac{C}{\sqrt{n}}\right)\exp\left[L_n\left(\hat{A}_n + \frac{C}{\sqrt{n}}\right) - L_n(\hat{A_n})\right]dC}
        \end{align*}
        Notice that 
        $$H = -\frac{1}{n} \frac{\partial^2}{\partial A\partial A}(\hat{A_n})\rightarrow I(A_0)\;\;\;a.s.\;P_{A_0}.$$
        It suffices to show that 
        \begin{equation}
        \begin{aligned}
        \int_{\R^{p\times p}} \left|J\left(\by, \hat{A}_n + \frac{C}{\sqrt{n}}\right) \exp\left[L_n\left(\hat{A}_n + \frac{C}{\sqrt{n}}\right) - L_n(\hat{A_n})\right] \right.&\\
        \left.- \pi_{A_0}(A_0)\exp\left[ \frac{-C^TI(A_0)C}{2}\right] \right| dC &\xrightarrow{P_{A_0}} 0
        \end{aligned}
        \label{eq:BvM1}
        \end{equation}
        
        Let $C_x$ be the $ij$th entry of C, where $x = i + (p-1)j$. By Taylor Theorem, 
        \begin{eqnarray*}
                L_n\left(\hat{A}_n + \frac{C}{\sqrt{n}}\right) &= & L_n(\hat{A}_n) + \sum_{x = 1}^{p^2}\left(\frac{C_x}{\sqrt{(n)}} \right) \frac{\partial}{\partial A_x}L_n(\hat{A}_n)\\
                && + \frac{1}{2}\sum_{x = 1}^{p^2}\sum_{y = 1}^{p^2}\left(\frac{C_xC_y}{(\sqrt{(n)})^2} \right) \frac{\partial^2}{\partial A_x\partial A_y}L_n(\hat{A}_n)\\
                && + \frac{1}{6}\sum_{x = 1}^{p^2}\sum_{y = 1}^{p^2}\sum_{z = 1}^{p^2}\left(\frac{C_xC_yC_z}{(\sqrt{(n)})^3} \right) \frac{\partial^3}{\partial A_x\partial A_y\partial A_z}L_n(A')\\
                &=&  L_n(\hat{A}_n) - \frac{C^THC}{2} + R_n
        \end{eqnarray*}
        for some $A' \in \left[\hat{A}_n, \hat{A}_n+\frac{C}{\sqrt{n}} \right]$. Notice that $R_n = \mathcal{O}p(n^{-3/2}||C||)$.
        Given any $0<\delta<\delta_0$ and $t>0$, the parameter space $\R^{p\times p}$ can be partitioned into three regions:
        \begin{align*}
        S_1 & = \{C: ||C|| < t\log\sqrt{n}\}\\
        S_2 & = \{C: t\log\sqrt{n}<||C|| < \delta\sqrt{n}\}\\
        S_3 & = \{C: ||C|| > \delta\sqrt{n}\}\\
        \end{align*}
        On $S_1\cup S_2$, 
        \begin{eqnarray*}
                &&\int_{S_1\cup S_2} \left|J\left(\by, \hat{A}_n + \frac{C}{\sqrt{n}}\right) \exp\left[L_n\left(\hat{A}_n + \frac{C}{\sqrt{n}}\right) - L_n(\hat{A}_n)\right] \right.\\
                &&\left.- \pi_{A_0}(A_0)\exp\left[ \frac{-C^TI(A_0)C}{2}\right] \right| dC\\
                &\leq& \int_{S_1\cup S_2} \left|J\left(\by, \hat{A}_n + \frac{C}{\sqrt{n}}\right) - \pi_{A_0}\left(\hat{A}_n + \frac{C}{\sqrt{n}}\right)\right|\\
                &&\times \exp\left[L_n\left(\hat{A}_n + \frac{C}{\sqrt{n}}\right) - L_n(\hat{A}_n)\right] dC\\
                &&+\int_{S_1\cup S_2} \left|\pi_{A_0}\left(\hat{A}_n + \frac{C}{\sqrt{n}}\right) \exp\left[L_n\left(\hat{A}_n + \frac{C}{\sqrt{n}}\right) - L_n(\hat{A}_n)\right] \right.\\
                &&\left.- \pi_{A_0}(A_0)\exp\left[ \frac{-C^TI(A_0)C}{2}\right] \right| dC
        \end{eqnarray*}
        Since $\pi_{A_0}(\cdot)$ is a proper prior on the region $S_1\cup S_2$, the second term goes to zero by the Bayesian Bernstein-von Mises Theorem (see the proof of Theorem 1.4.2 in \cite{Ghosh2003}). \\
        Next we notice that 
        \begin{eqnarray*}
                &&\int_{S_1\cup S_2} \left|J\left(\by, \hat{A}_n + \frac{C}{\sqrt{n}}\right) - \pi_{A_0}\left(\hat{A}_n + \frac{C}{\sqrt{n}}\right)\right|\\
                &&\times \exp\left[L_n\left(\hat{A}_n + \frac{C}{\sqrt{n}}\right) - L_n(\hat{A}_n)\right] dC\\
                &\leq & \sup_{C\in S_1\cup S_2}  \left|J\left(\by, \hat{A}_n + \frac{C}{\sqrt{n}}\right) - \pi_{A_0}\left(\hat{A}_n + \frac{C}{\sqrt{n}}\right)\right|\\
                && \times\int_{S_1\cup S_2} \exp\left[L_n\left(\hat{A}_n + \frac{C}{\sqrt{n}}\right) - L_n(\hat{A}_n)\right] dC
        \end{eqnarray*} 
        Since $\sqrt{n}\left( \hat{A}_n - A_0\right) \xrightarrow{\mathcal{D}} N\left(0, I(A_0)^{-1}\right)$, we have 
        $$P_{A_0}\left[\left\{\hat{A}_n + \frac{C}{\sqrt{n}}; \; C\in S_1\cup S_2\right\}\subset B(A_0,\delta_0)\right]\rightarrow 1.$$
        Furthermore, 
        $$L_n\left(\hat{A}_n + \frac{C}{\sqrt{n}}\right) - L_n\left(\hat{A}_n\right) = -\frac{C^THC}{2} + R_n,$$
        so the integral converges in probability to 1. Since $\max_{C\in S_1\cup S_2} \leq \delta$ and $J_n \rightarrow \pi_{A_0}$, the term goes to 0 in probability. \\
        Turning our attention to $S_3$, notice that
        \begin{eqnarray*}
                &&\int_{S_3} \left|J\left(\by, \hat{A}_n + \frac{C}{\sqrt{n}}\right) \exp\left[L_n\left(\hat{A}_n + \frac{C}{\sqrt{n}}\right) - L_n(\hat{A}_n)\right] \right.\\
                &&\left.- \pi_{A_0}(A_0)\exp\left[ \frac{-C^TI(A_0)C}{2}\right] \right| dC\\
                &\leq& \int_{S_3}J\left(\by, \hat{A}_n + \frac{C}{\sqrt{n}}\right)\exp\left[L_n\left(\hat{A}_n + \frac{C}{\sqrt{n}}\right) - L_n(\hat{A}_n)\right] dC\\
                &&+ \int_{S_3} \pi_{A_0}(A_0)\exp\left[ \frac{-C^TI(A_0)C}{2}\right]dC
        \end{eqnarray*}
        The last integral goes to zero in $P_{A_0}$ because $\min_{S_3}||C|| \rightarrow \infty$. As for the first integral, under the $l_\infty$-norm,

        \begin{eqnarray*}
                && \int_{S_3}J\left(\by, \hat{A}_n + \frac{C}{\sqrt{n}}\right)\exp\left[L_n\left(\hat{A}_n + \frac{C}{\sqrt{n}}\right) - L_n(\hat{A}_n)\right] dC\\
                & = &\begin{pmatrix} n\\p \end{pmatrix}^{-1} \sum_{\i} \int_{S_3}J\left(\by_{\i}, \hat{A}_n + \frac{C}{\sqrt{n}}\right)\exp\left[L_n\left(\hat{A}_n + \frac{C}{\sqrt{n}}\right) - L_n(\hat{A}_n)\right] dC\\
                & = & \begin{pmatrix} n\\p \end{pmatrix}^{-1} \sum_{\i} \int_{S_3}J\left(\by_{\i}, \hat{A}_n + \frac{C}{\sqrt{n}}\right)f\left(\by_{\i}| \hat{A}_n + \frac{C}{\sqrt{n}}\right)\\
                &&\times\exp\left[L_n\left(\hat{A}_n + \frac{C}{\sqrt{n}}\right) - L_n(\hat{A}_n)- \log f\left(\by_{\i}| \hat{A}_n + \frac{C}{\sqrt{n}}\right) \right] dC\\
        \end{eqnarray*}
        
        Under the $l_\infty$-norm, $\int J(\by_{\i}, A)f(\by_{\i}, A)dA = 1,\; \forall \i$; Proposition \ref{prop:minLA} guarantees that the exponent goes to $-\infty$. Thus, the integral goes to zero in probability. 

        Under the $l_2$-norm, for each $\by$, let $\i$ be 
        $$\i = \argmin_{\tilde{\i}} \left|J\left(\by, \hat{A}_n + \frac{C}{\sqrt{n}}\right) - J\left(\by_{\tilde\i}, \hat{A}_n + \frac{C}{\sqrt{n}}\right)\right| = \argmin_{\tilde{\i}} h(\by, C, \tilde{\i}).$$
        \begin{eqnarray*}
                && \int_{S_3}J\left(\by, \hat{A}_n + \frac{C}{\sqrt{n}}\right)\exp\left[L_n\left(\hat{A}_n + \frac{C}{\sqrt{n}}\right) - L_n(\hat{A}_n)\right] dC\\
                & \leq & \int_{S_3}\left\{h(\by, C, \i)f\left(\by_{\i}| \hat{A}_n + \frac{C}{\sqrt{n}}\right) + J\left(\by_{\i}, \hat{A}_n + \frac{C}{\sqrt{n}}\right)f\left(\by_{\i}| \hat{A}_n + \frac{C}{\sqrt{n}}\right) \right\}\\
                && \exp\left[L_n\left(\hat{A}_n + \frac{C}{\sqrt{n}}\right) - L_n(\hat{A}_n)- \log f\left(\by_{\i}| \hat{A}_n + \frac{C}{\sqrt{n}}\right) \right] dC\\
        \end{eqnarray*}
        Note that as $n$ goes to infinity, the first two product terms, $h(\cdot)f(\cdot)$ and $J(\cdot)f(\cdot)$, are both bounded; the exponent term goes to $-\infty$ by Proposition \ref{prop:minLA}, so the integral goes to zero in probability.
        
        Having shown Eq \ref{eq:BvM1}, we now follow Ghosh and Ramamoorthi \cite{Ghosh2003} and let 
        $$D_n = \int_{R^{p\times p}} \left|J\left(\by, \hat{A}_n + \frac{C}{\sqrt{n}}\right) \exp\left[L_n\left(\hat{A}_n + \frac{C}{\sqrt{n}}\right) - L_n(\hat{A}_n)\right] \right| dC$$
        Then the main result to be proven (Eq \ref{eq:BvM0}) becomes 
        \begin{equation}
        \begin{aligned}
        & D_n^{-1}\left\{\int_{\R^{p\times p}}\left|J\left(\by, \hat{A}_n + \frac{B}{\sqrt{n}}\right) \exp\left[L_n\left(\hat{A}_n + \frac{B}{\sqrt{n}}\right) - L_n(\hat{A}_n)\right] \right.\right.\\
        &\left.\left.   -D_n\frac{\sqrt{\det(I(A_0))}}{\sqrt{2\pi}}\exp\left(-\frac{B^TI(A_0)B}{2}      \right)  \right|        \right\}dB \xrightarrow{P_{A_0}} 0
        \end{aligned}
        \label{eq:BvM2}
        \end{equation}
        Because
        \begin{eqnarray*}
                && \int_{\R^{p\times p }} J(\by,\hat{A}_n)\exp\left(-\frac{B^TI(A_0)B}{2}\right)        dB\\
                & = & J(\by,\hat{A}_n) \int_{\R^{p\times p }} \exp\left(-\frac{B^TI(A_0)B}{2}\right)    dB\\
                & = & J(\by,\hat{A}_n) \frac{\sqrt{2\pi}}{\sqrt{\det(H)}}\\
                &\xrightarrow{a.s.}& \pi(A_0)\frac{\sqrt{2\pi}}{\sqrt{\det(H)}}
        \end{eqnarray*}
        and Eq \ref{eq:BvM1} implies that $D_n\xrightarrow{P}  \pi(A_0)\frac{\sqrt{2\pi}}{\sqrt{\det(H)}}$. It is sufficient to show that the integral in Eq \ref{eq:BvM2} goes to 0 in probability. This integral is less than $I_1 + I_2$, where 
        \begin{eqnarray*}
                I_1 & = & \int_{\R^{p\times p}}\left|J\left(\by, \hat{A}_n + \frac{B}{\sqrt{n}}\right) \exp\left[L_n\left(\hat{A}_n + \frac{B}{\sqrt{n}}\right) - L_n(\hat{A}_n)\right] \right.\\
                && \left. -J\left(\by, \hat{A}_n\right)\exp\left(-\frac{B^TI(A_0)B}{2}  \right)  \right|dB
        \end{eqnarray*}
        and
        \begin{eqnarray*}
                I_2 & = & \int_{\R^{p\times p}}\left|J\left(\by, \hat{A}_n\right)\exp\left(-\frac{B^THB}{2}     \right)\right.\\
                && \left.-D_n\frac{\sqrt{\det(I(A_0))}}{\sqrt{2\pi}}\exp\left(-\frac{B^TI(A_0)B}{2}\right) \right|dB
        \end{eqnarray*}
        Eq \ref{eq:BvM1} shows that $I_1\rightarrow 0$ in probability.\\
        Since
        \begin{align*}
        & J(\by, \hat{A}_n)  \xrightarrow{P} \pi(A_0)\\
        & D_n \xrightarrow{P} \pi(A_0)\frac{\sqrt{2\pi}}{\sqrt{\det(I(A_0))}}
        \end{align*}
        we have 
        \begin{eqnarray*}
                I_2 &=& \left|J\left(\by, \hat{A}_n\right) - D_n\frac{\sqrt{\det(I(A_0))}}{\sqrt{2\pi}}\right| \int_{\R^{p\times p}}\exp\left(-\frac{B^THB}{2}  \right)dB\xrightarrow{P}0
        \end{eqnarray*}
\end{proof}

\subsection{Derivation of the normalization constant Eq (\ref{eq:NormalizingConstant})}\label{sec:normalizingconstant}

Using a substitution $A^{-1}(nS_n)^{1/2}=\bZ$ with the Jacobian
$dA=|\det \bZ|^{-2p} |\det (nS_n)|^{p/2}dZ$ we have
\begin{align*}
\int J(\by,A)f(\by|A)\,dA&=
C(\by)\int\frac{e^{-\frac 12\operatorname{tr} (A^{-1}(n S_n)^{1/2}) (A^{-1}(n S_n)^{1/2})^\top}}{(2\pi)^{np/2} |\det A|^{n+p}}\,dA\\
&=C(\by)\int |\det \bZ|^{n-p}|\det (n S_n)|^{-n/2}e^{-\frac 12\operatorname{tr} \bZ\bZ^\top}\,d\bZ\\
&=(2\pi)^{-(n-p)p/2}C(\by)|\det (n S_n)|^{-n/2} E |\det \bZ|^{n-p}\\
&=\frac{\pi^{(p^2 - np)/2} C(\by)\Gamma_p\left(\frac n2\right)}{|\det (nS_n)|^{n/2}\Gamma_p\left(\frac p2\right)}
\end{align*}
The last equality follows from the fact that for a $p\times p$ matrix of independent standard normal normal variables $Z$ we have
\[
E |\det \bZ|^{n} = \frac{2^{np/2} \Gamma_p\left(\frac{n+p}{2}\right)}{\Gamma_p\left(\frac p2\right)}.
\]

\subsection{Proof of Lemma \ref{l:constant}}

\begin{lemma*}[\ref{l:constant}]
        For any clique model $\cM$ with $k$ cliques of sizes $g_i,\ i=1,\ldots k$ we have 
        \begin{itemize}
                \item[(1)] under the $l_2$ norm, $C_{\cM,i}(\by)=|\det S_{n}^{\cM, i}|^{g_i/2} $,
                $$C_{\cM,i}(\by)
                \to
                |\det (\Sigma_0^{\cM,i})|^{\frac{g_i}{2}};
                $$
                
                \item[(2)] under the $l_\infty$ norm, 
                $C_{\cM,i}(\by) = \begin{pmatrix}
                n\\g_i
                \end{pmatrix}^{-g_i}\left(\Jsumgi\left|\det\left(V_{\cM,i}\right)_{\i_i} \right|\right)^{g_i},$
                $$C_{\cM,i}(\by)
                \to 
                |\det (\Sigma_0^{\cM,i})|^{\frac{g_i}{2}} 2^{\frac{g_i^2}{2}}\pi^{-\frac{g_i}{2}}\Gamma\left(\frac{g_i + 1}{2}\right)^{g_i},\; \text{a.s.}$$ 
                where $S_{n}^{\cM, i}$ is the sample covariance computed using only observations within clique $i$, $V_{\cM,i}$ is the sub-matrix of $V$ that only includes the observations in clique $i$, and $\Sigma_0^{\cM,i}$ denotes the $i$th block component of $ \Sigma_0^{\cM}$.              
        \end{itemize}   
\end{lemma*}

\begin{proof}
        When dealing with the $l_2$-norm, the Strong Law of Large Numbers implies $ S_{n,i}^\cM\to\Sigma_0^{\cM,i}$ a.s. for each $i=1, \ldots, k$ and the results follows by continuity.
        
        In the case of the $l_\infty$ norm   the convergence theorem for U-statistics gives 
        $C_i(\by)\to E |\det V|^{g_i}=\left(\Sigma_0^{\cM,i}\right)^{g_i/2}  \left(E|\det  \bZ_i|\right)^{g_i}$ a.s. Here $\bZ_i$ denotes a $g_i\times g_i$ matrix of standard Gaussian random variables. Since $E|\det  \bZ_i|=2^{g_i/2}\pi^{-1/2}\Gamma\left(\frac{g_i+1}2\right)$ the result follows \citep{McLennan2002}.
        
\end{proof}

\subsection{Proof of Lemma \ref{l:gamma}}
\begin{lemma*}[\ref{l:gamma}]
        Let $g_i, i=1,\ldots, k$ and $h_j, j=1,\ldots,l$ be integers such that $\sum_{i=1}^k g_i=\sum_{j=1}^l h_i$. Then
        as $n\to\infty$
        \[
        \frac{\prod_{i=1}^k \Gamma_{g_i}\left(\frac{n}{2}\right)}{\prod_{j=1}^l \Gamma_{h_j}\left(\frac{n}{2}\right)}
        \sim
        \left(\frac{\pi}n\right)^{\frac{\sum_{i=1}^k g_i^2-\sum_{j=1}^k h_j^2}4}.
        \]
\end{lemma*}
\begin{proof}
        It is the well-known \citep{Abramowitz1964} that
        \begin{equation}\label{eq:gammaratio}
        \frac{\Gamma(x+y)}{\Gamma(x)}\sim x^y,\quad\mbox{as $x\to\infty$ and $y$ is fixed.}
        \end{equation}
        Recall
        \[
        \frac{\prod_{i=1}^k \Gamma_{g_i}\left(\frac{n}{2}\right)}{\prod_{j=1}^l \Gamma_{h_j}\left(\frac{n}{2}\right)}
        =\frac{\pi^{\sum_{i=1}^k (g_i^2-g_i)/4}\prod_{i=1}^k \prod_{s=1}^{g_i} \Gamma\left(\frac{n+1-s}{2}\right)}
        {\pi^{\sum_{i=1}^l (h_i^2-h_i)/4}\prod_{j=1}^l \prod_{t=1}^{h_j} \Gamma\left(\frac{n+1-t}{2}\right)}
        \]
        Since both numerator on denominator includes a product of $p$ gamma functions, the result of the lemma then follows directly from Eq \ref{eq:gammaratio}. Note that Eq \ref{eq:gammaratio} will be sufficient when $p$ is fixed. More precise bounds available in \cite{Jameson2013} could be used when $p$ is growing with $n$.
        
\end{proof}

\subsection{Proof of Lemma \ref{l:detSnratio}}

\begin{lemma*}[\ref{l:detSnratio}]
        Let $\cM$ be a clique model. 
        \begin{itemize}
                \item[i.] If $\det(\Sigma_0)<\det(\Sigma_0^\cM)$, then there is $a>0$, such that
                \[
                \left|\frac{\det (S_n^{\cM_0})}{\det (S_n^\cM)}\right|^{n/2} \leq e^{-an}\quad\mbox{eventually a.s.}
                \]
                
                \item[ii.] If $\cM\neq\cM_0$ is compatible with $\Si_0$, then as $n\to\infty$
                \[
                \left|\frac{\det (S_n^{\cM_0})}{\det (S_n^\cM)}\right|^{n/2} = \mathcal{O}_P(1).
                \]
        \end{itemize}
\end{lemma*}

\begin{proof}
        If  $\det(\Sigma_0)<\det(\Sigma_0^\cM)$, set 
        $
        a = \frac{\log\det\Sigma_0^\cM-\log\det\Sigma_0}{4}.
        $
        By the Strong Law of Large Numbers, 
        \[
        S_n^{\cM_0}\to\Sigma_0,\;\;\;S_n^\cM\to\Sigma_0^\cM, \text{ a.s.}
        \]
        Thus eventually a.s. 
        ${\det S_n^{\cM_0}}/{\det S_n^\cM}<e^{-a}$ and the statement of the lemma follows.
        
        If $\cM\neq\cM_0$ is compatible with $\Si_0$, by the Central Limit Theorem 
        \[
        \sqrt{n} (S_n^\cM - S_n^{\cM_0})\toD R.
        \]
        By Slutsky's theorem the spectral radius and minimum eigenvalue of $(S_n^{\cM_0})^{-1}(S_n^\cM-S_n^{\cM_0})$ satisfy
        $\rho=\mathcal{O}_P(n^{-1/2})$ and $\lambda=o_P(1)$ respectively. Consequently by \ref{eq:spectralbound}
        \[
        \left|\frac{\det S_n^{\cM_0}}{\det S_n^\cM}\right|^{n/2}\leq e^{\frac{np\rho^2}{2(1+\lambda)}}=\mathcal{O}_P(1).
        \]
\end{proof}

\subsection{Proof of Theorem \ref{t:clique}}

\begin{thm*}[\ref{t:clique}]
        For any clique model $\cM$ that is not compatible with $\Sigma_0$  assume $\det\left(\Sigma_0\right)<\det\left(\Sigma_0^\cM\right)$ and  the penalty  $e^{-a n} q_\cM(n)/q_{\cM_0}(n)\to 0$ for all $a>0$ as $n\to 0$.
        
        For any clique model $\cM$ compatible with $\Sigma_0$  assume that 
        $q_\cM(n)/q_{\cM_0}(n)$ is bounded.
        
        Then as $n\to\infty$ with $p$ held fixed
        $
        r_p(\cM_0|\bY)\toP 1. 
        $
\end{thm*}

\begin{proof}
        Because for any fixed $p$ there are  finitely many clique models, we only need to prove that for any $\cM\neq\cM_0$,
        $\frac{r_p(\cM|\bY)}{r_p(\cM_0|\bY)}\toP 0$.   
        
        Denote by $g_i, i=1,\ldots,k$, the size of cliques in $\cM$ and 
        $h_j, j=1,\ldots,l$, the size of cliques in $\cM_0$.
        
        By Lemma~(\ref{l:constant}), (\ref{l:gamma}) we have as $n\to\infty$
        \[
        \frac{r_p(\cM|\bY)}{r_p(\cM_0|\bY)}  \sim Kn^{-\frac{\sum_{i=1}^k g_i^2-\sum_{j=1}^l h_j^2}4} 
        \frac{q_\cM(n)}{q_{\cM_0}(n)}  \left|\frac{\det S_n^{\cM_0}}{\det S_n^\cM}\right|^{n/2},
        \]
        where $K$ is a constant independent of $n$.
        
        If $\cM$ is not compatible with $\Sigma_0$ by assumption and Lemma~(\ref{l:detSnratio}.$i$) we have $\frac{r_p(\cM|\bY)}{r_p(\cM_0|\bY)}\to 0$ a.s. 
        
        If $\cM\neq\cM_0$ is compatible with $\Sigma_0$ notice that $\cM$ is obtained by pooling together some cliques of $\cM_0$. Therefore $\sum_{i=1}^k g_i^2-\sum_{j=1}^l h_j^2>4$. Consequently $\frac{r_p(\cM|\bY)}{r_p(\cM_0|\bY)}\toP 0$ by assumption and Lemma~\ref{l:detSnratio}.$ii$.
\end{proof}

\bibliographystyle{abbrv}
\bibliography{myProposal}

\end{document}


\begin{frontmatter}
		\title{Covariance estimation via fiducial inference \\ Supplementary: Additional Simulations}
		\runtitle{Covariance estimation via fiducial inference}
		
\begin{aug}
	\author{\fnms{W. Jenny} \snm{Shi}\thanksref{t1}\corref{}\ead[label=e1]{wjennyshi@gmail.com}}
	\address{Computational Bioscience Program, University of Colorado
		\printead{e1}}
	\author{\fnms{Jan} \snm{Hannig}\thanksref{t2}
		\ead[label=e2]{jan.hannig@unc.edu}}
	\address{Department of Statistics \& Operations Research, University of North Carolina
		\printead{e2}}
	\author{\fnms{Randy C.S.} \snm{Lai}
		\ead[label=e3]{chushing.lai@maine.edu}}
	\address{Department of Mathematics \& Statistics, University of Maine
		\printead{e3}}
	\author{\fnms{Thomas C.M.} \snm{Lee}\thanksref{t3}
		\ead[label=e4]{tcmlee@ucdavis.edu}}
	\address{Department of Statistics, University of California
		\printead{e4}}

	\thankstext{t1}{Shi's research was supported in part by the National Library of Medicine Institutional Training Grant T15LM009451}
	\thankstext{t2}{Hannig's research was supported in part by the National Science Foundation (NSF) under Grant  No.\ 1512945 and 1633074}
	\thankstext{t3}{Lee's research was supported in part by the NSF under Grant  No.\ 1512945 and 1513484}
	\runauthor{Shi, Hannig, Lai, Lee}

\end{aug}

	\end{frontmatter}

%
%
	

		\section{Clique example with $\Sigma: \; k =8, p = 100, n = 1000$}
		
\begin{SCfigure}[][h]
			\includegraphics[height = 9.5cm]{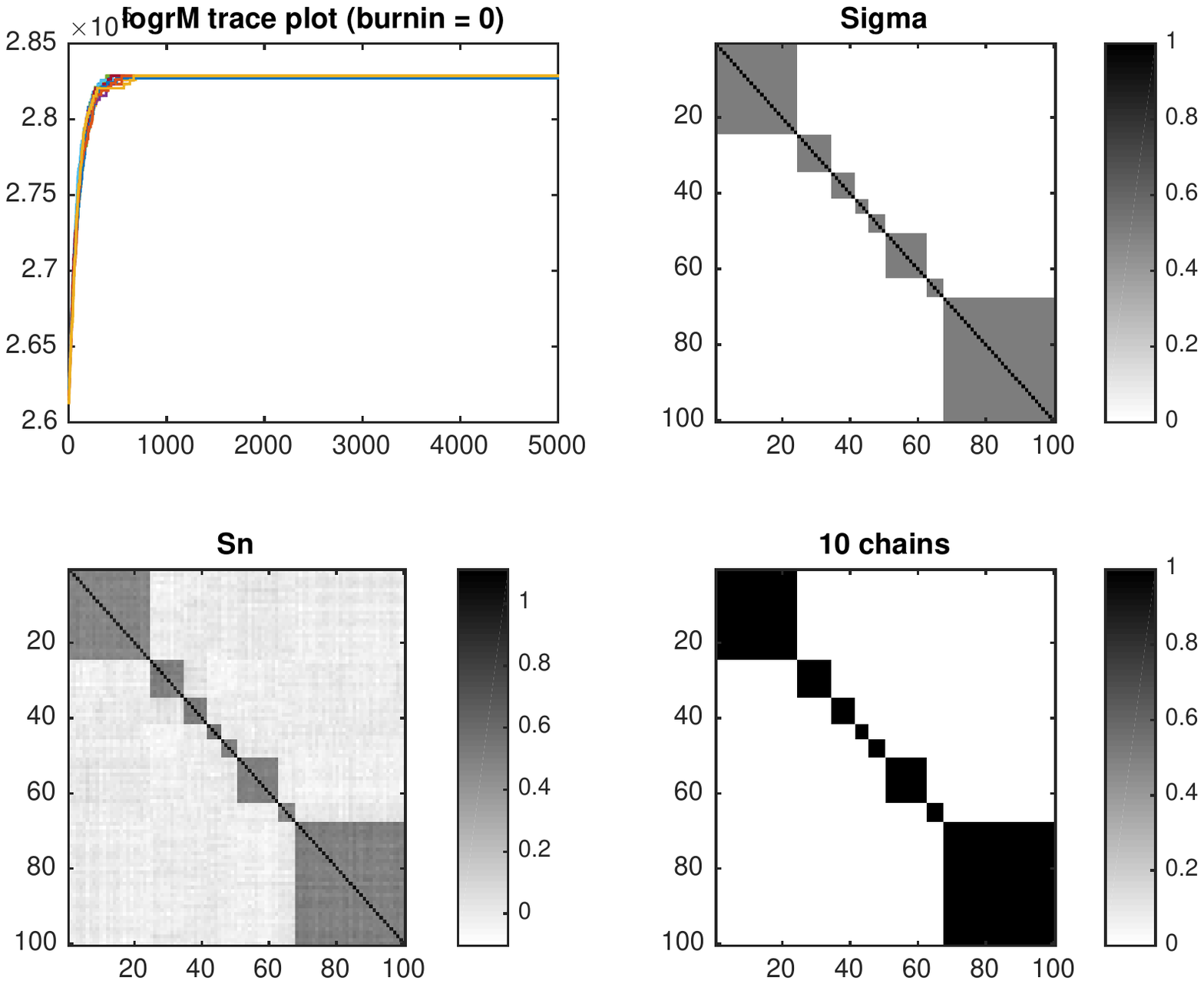}
			\vspace{-.9in}
			\caption{\label{fig:Cliquep100n1000} Result for $k =8, p = 100, n = 1000$. The trace plot (top left) shows that the chains converge quickly. Although$\frac{n}{p}$ is small, the sample covariance (bottom left) roughly captures the shape of true covariance (top right). The last panel (bottom right) shows that the fiducial estimate captures the true clique structure perfectly. }
		\end{SCfigure}

\begin{SCfigure}[][h]
			\includegraphics[height = 9.5cm]{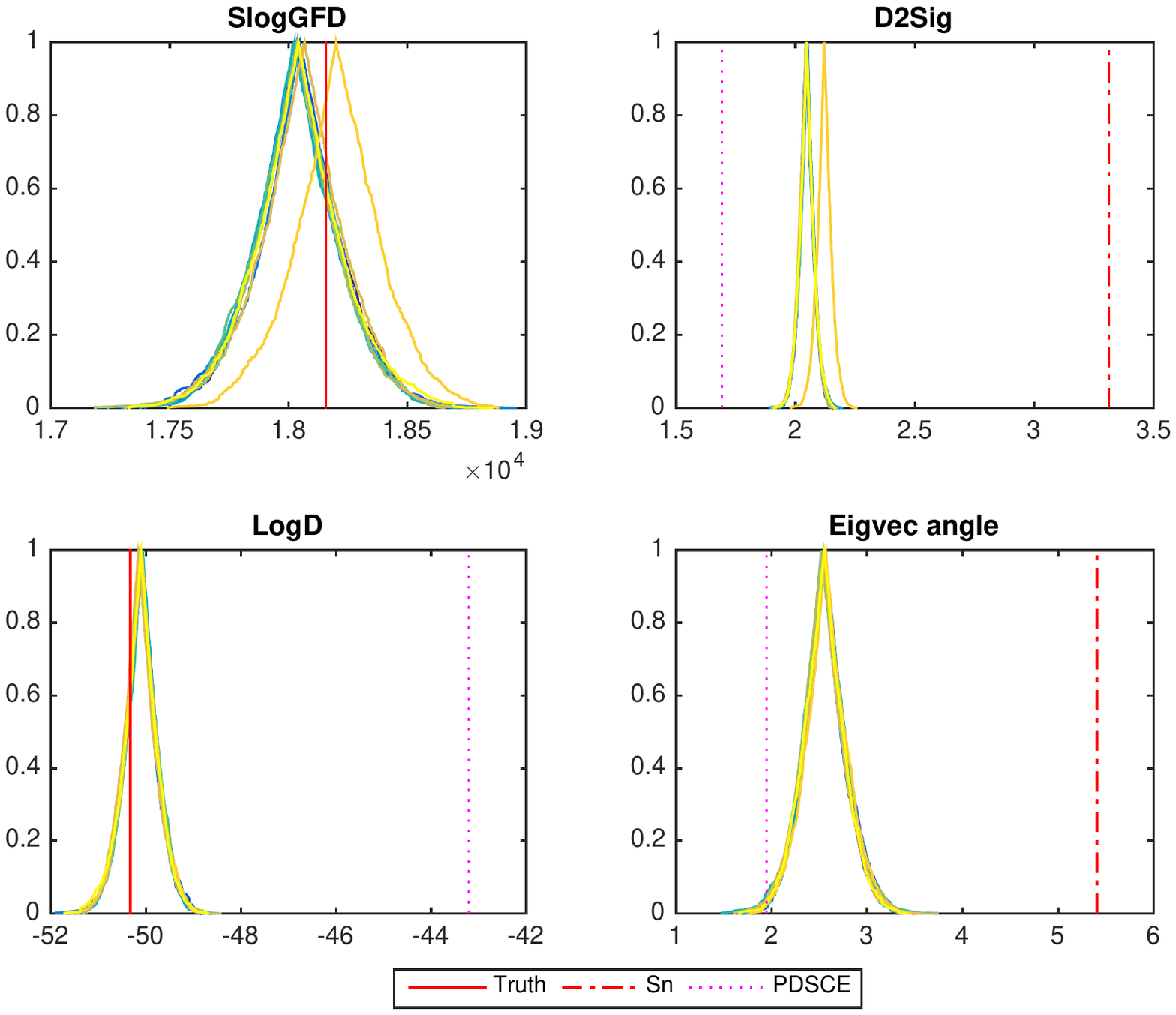}
			\vspace{-.9in}
			\caption{\label{fig:Cliquep100n1000cc} Confidence curve plots for estimated covariance matrix. $k =8, p = 100, n = 1000$. Comparing to the sample covariance, the estimators are closer to $\Sigma$. The PDSCE estimator shows even smaller FM-distance to $\Sigma$, it, however, greatly overestimates $\det{\Sigma}$. }
		\end{SCfigure}

		Figure \ref{fig:Cliquep100n1000} shows the sparse structure estimation for a simulation with 10 chains started at random. The bottom right panel is based on the last 1000 iterations. The estimated cliques match true covariance exactly with fiducial probability one. Its corresponding confidence curve plots, along with the PDSCE estimator is shown in Figure \ref{fig:Cliquep100n1000cc}. Left two panels show that the estimators peak very close to the truth. The right two panels indicate that the fiducial estimators are better than the sample covariance. Similar to previous examples, the PDSCE shows smaller distance to the true covariance matrix, but over estimates the log-determinant. 
		
\begin{SCfigure}[][h]
			\includegraphics[height = 9.5cm]{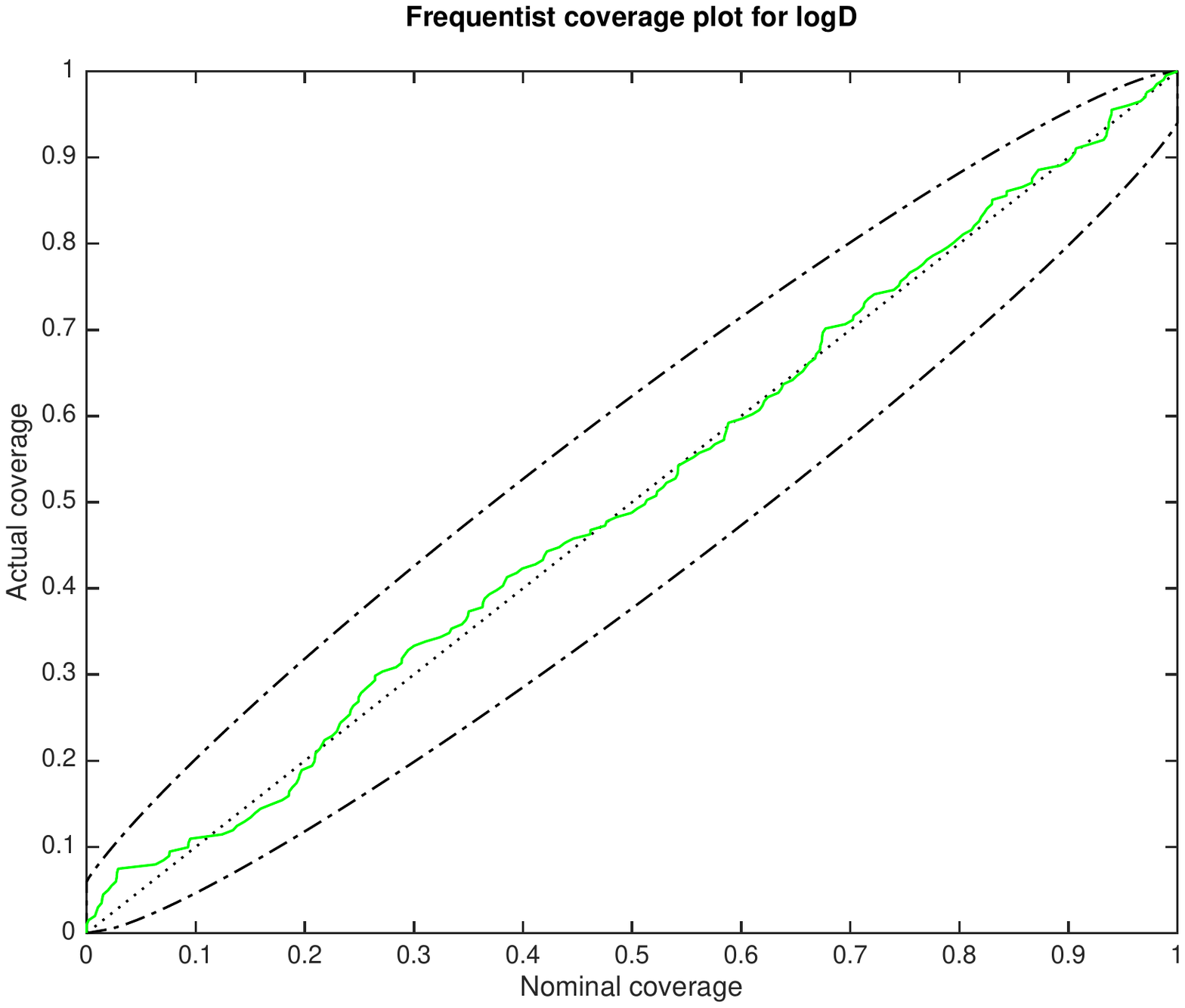}
			\vspace{-.9in}
			\caption{\label{fig:Cliquep100n1000logD} 95\% coverage log-determinant plots for 200 repeated simulations. $k =8, p = 100, n = 1000$. The p-values (in green) roughly follow a uniform [0,1] distribution. It shows well calibrated coverage.  }
		\end{SCfigure}

		Repeat the simulation 200 times, each with one MCMC, we obtain the one-sided p-values. Its quantile-quantile plot against Unif[0,1] is shown in Figure \ref{fig:Cliquep100n1000logD}. The p-value curve (in green) weaves around the diagonal dotted guideline. It is well contained by the 95\% coverage envelope, indicating good coverage. 
		
		\section{ $\Sigma$ with zero locations known:$\; p = 50, n = 50$}
		
\begin{SCfigure}[][h]
			\includegraphics[height = 9.5cm]{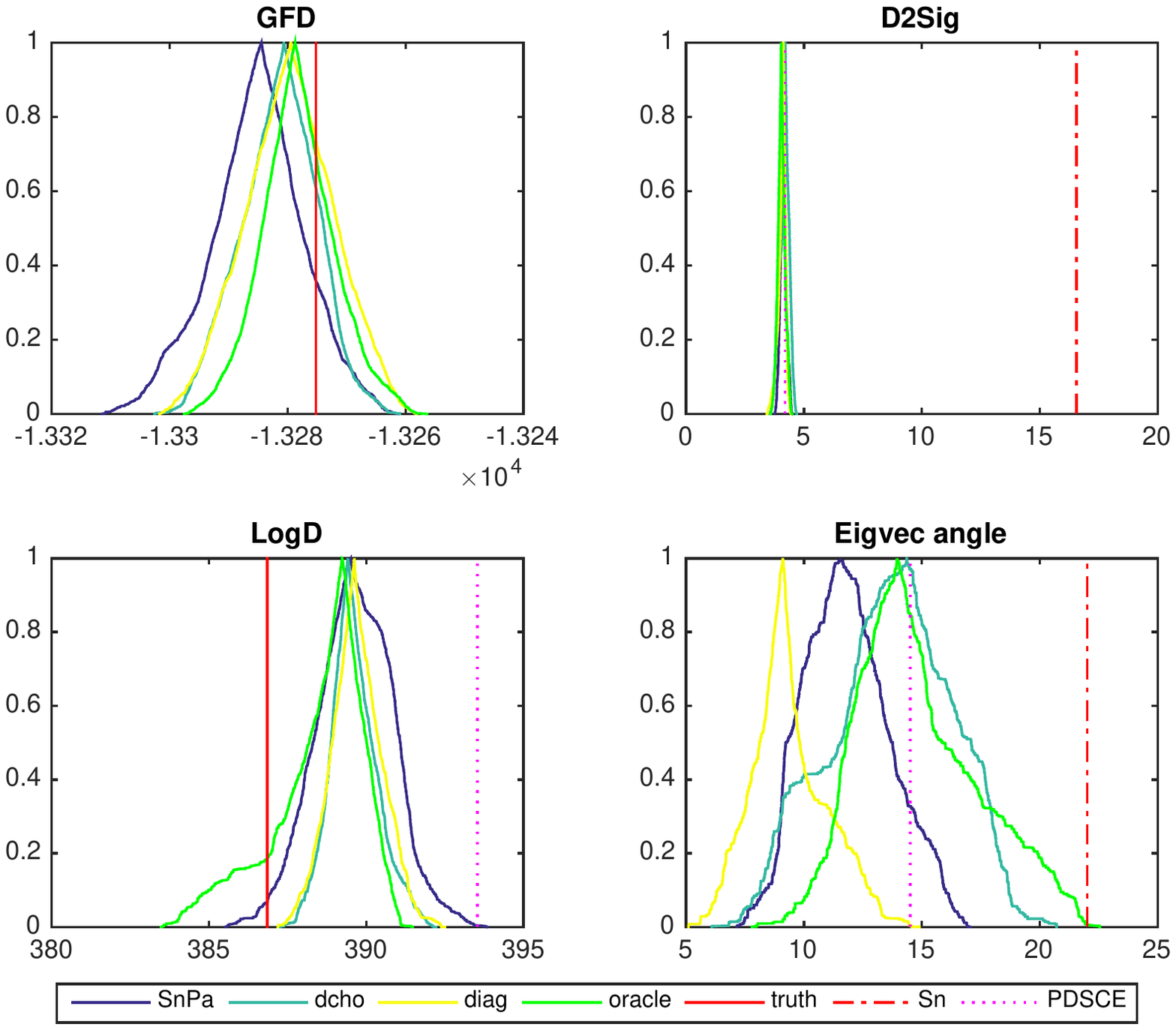}
			\vspace{-.9in}
			\caption{\label{fig:MCMCp50n50} Overall, the fiducial estimators are slightly better than the PDSCE estimator in this example. They are both closer to the truth than the sample covariance matrix.}
		\end{SCfigure}

		The confidence curves peak near the truth in the GFD panel, while slightly overestimates the covariance determinant. The PDSCE estimator overestimates the covariance determinant again. Our method and PDSCE produce estimators that are closer to the truth than the sample covariance. The last panel shows that the fiducial estimators have a relatively wide range of angles, yet the peaks are at least as good as the PDSCE estimator. For each Markov chain the burn in = 1000, and the window = 5000 (Figure \ref{fig:MCMCp50n50}). 
		
		\section{General case: $ p = 50, n = 50, maxC = 5$ (fixed zero locations unknown)}
\begin{SCfigure}[][h]
			\includegraphics[height = 9.5cm]{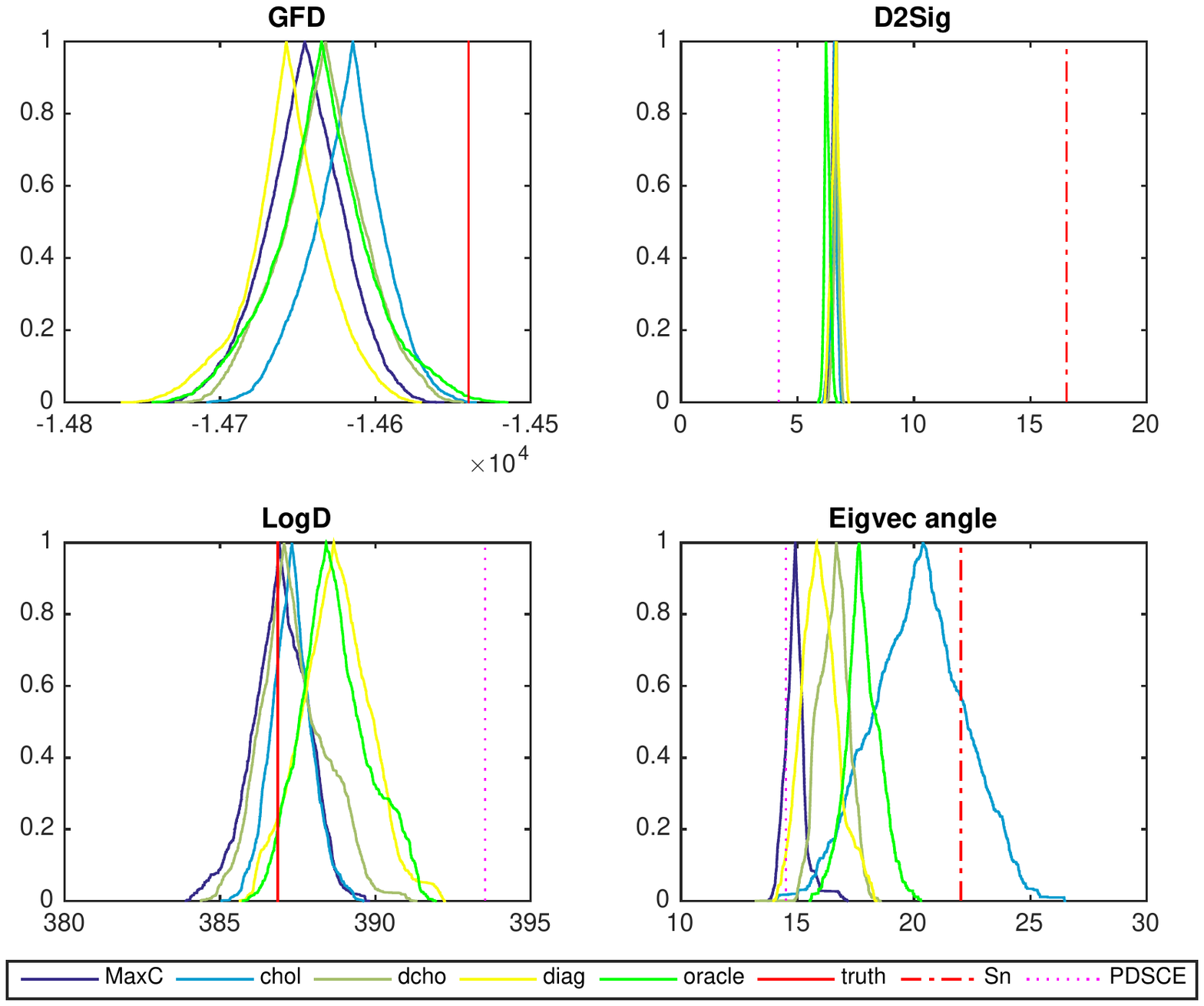}
			\vspace{-.9in} 
			\caption{\label{fig:RJp50n50} The fiducial estimators are better than the sample covariance matrix in the D2Sig and the Eigvec angle panels. The PDSCE estimator seems to be slightly better in those two panels, but overestimates the covariance determinant.  }
		\end{SCfigure}

		Working with the same true covariance matrix and observations as Example A2, we implement the algorithm for the general case where the sparse structure is unknown. The estimators fiducial are better than the sample covariance matrix. The PDSCE estimator appears to be slightly closer to the truth in FM-distance. Burn in = 50000, window = 10000 (See Figure \ref{fig:RJp50n50}).\\
		
